\documentclass[11pt,graphicx,amsmath]{article}
\usepackage{amsmath}
\usepackage{graphicx}
\usepackage{bm}
\usepackage[dvips]{color}
\usepackage{amssymb}
\usepackage{amsfonts}
\usepackage{comment}
\usepackage{cite}
\usepackage{todonotes}
\usepackage{caption}
\usepackage{subcaption}

\def\be{\begin{equation}}
\def\ee{\end{equation}}
\def\nn{\nonumber}

\def\ba{\begin{eqnarray}}
\def\ea{\end{eqnarray}}
\def\bl#1\el{\begin{align}#1\end{align}}

\def\l{\left}
\def\r{\right}

\title{  Maxwell field with gauge fixing term in de Sitter space:
              exact solution  and  stress tensor }
\author{\small
             Yang  Zhang\thanks{yzh@ustc.edu.cn} \,  and
           Xuan Ye   \thanks{yyyyy@mail.ustc.edu.cn}
              \\
 \small  Department of  Astronomy,
         CAS Key Laboratory for Researches in Galaxies and Cosmology, \\
 \small  School of Astronomy and Space Sciences, \\
 \small  University of Science and Technology of China,
              Hefei, Anhui, 230026, China \\
 }

 \date{}

\def\be{\begin{equation}}
\def\ee{\end{equation}}
\def\nn{\nonumber}

\allowdisplaybreaks

\large

\begin{document}

\maketitle

\begin{abstract}
\large

The Maxwell field with a general gauge fixing (GF) term is nontrivial,
not only the longitudinal and temporal modes are mixed up
in the field equations,
but also  unwanted consequences
might arise from the GF  term.
We derive the complete set of solutions in de Sitter space,
and implement the covariant canonical quantization
which restricts the residual gauge transformation
down to a quantum residual gauge transformation.
Then, in the Gupta-Bleuler (GB) physical state,
we calculate the stress tensor
which is amazingly independent of the gauge fixing constant
and is also invariant under the quantum residual gauge transformation.

The transverse components are simply
the same as those in the Minkowski spacetime,
and the transverse vacuum stress tensor
has only one   UV divergent term ($\propto  k^4$),
which becomes zero by the 0th-order adiabatic regularization.
The longitudinal-temporal stress tensor in the GB state
is zero due to a cancelation between
the longitudinal and temporal parts.
More interesting is the stress tensor of the GF term.
Its particle contribution is zero due to the cancelation in the GB state,
and its vacuum contribution is twice
that of a minimally-coupling massless scalar field,
containing $k^4$ and $k^2$ divergences.
After the 2nd-order adiabatic regularization,
the GF vacuum stress tensor becomes zero too,
so that there is no need to introduce a ghost field,
and the zero GF vacuum  stress tensor can not
be a possible candidate for the cosmological constant.
Thus,  all the physics
predicted by the Maxwell field  with the GF term
will be the same as that without the GF term.
We also carry out analogous calculation in the Minkowski spacetime,
and the stress tensor is
similar to, but simpler than that in de Sitter space.

\end{abstract}

\


      Quantum fields  in curved spacetime,
      Maxwell field,
      inflationary universe,

\large

\section{Introduction}

The Maxwell field
is well studied as a quantum field in the flat spacetime.
The canonical quantization is simple in radiation gauge
in which the temporal and  longitudinal components are set   zero,
only two transverse polarizations remain as dynamical variables.
However, when a general gauge fixing (GF) term is introduced for
a  covariant canonical quantization,
the longitudinal and temporal modes, $A$ and $A_0$,
are mixed up in their field equations,
and the solutions are nontrivial except in the Feynman gauge.
In curved spacetimes
the mixing-up of $A$ and $A_0$ occurs even  in  the Feynman gauge.
More seriously,
the introduced GF term  gives rise to
a part of stress tensor,
which would bring about unwanted consequences.
Conventionally, a ghost field is introduced
\cite{AdlerLiebermanNg1977,ChuKoyama2017}
to cancel out the GF stress tensor,
so that the net result will be  a zero stress tensor,
and no  unphysical consequence will occur.
In another treatment \cite{JimenezMaroto2009,JimenezMaroto2010},
the GF term was used
to play a role of the cosmological constant.
However, the  vacuum GF stress tensor is UV divergent,
and must be regularized
before considering its possible physical implication.
Ref.\cite{Vollick2012} adopted
the Dirac's approach to constrained system
to study the Maxwell field (without the GF term)
in a general RW spacetime,
and calculated the Hamiltonian.
But the  non-covariant Hamiltonian
is not the same as the stress tensor,
and the UV divergences and regularization were not addressed either.

In this paper,
we shall derive the complete set of  solutions of
the Maxwell field with a general GF term in de Sitter space,
and reveal the interesting structure of the solutions.
With these,  we shall implement the covariant canonical quantization,
and obtain its constraint on the coefficients of solution modes,
as well as its restriction on the residual gauge transformation.
Then we shall calculate respectively the transverse stress tensor,
the longitudinal and temporal stress tensor in the Gupta-Bleuler (GB) state,
and the stress tensor due to the GF term
in the GB physical state \cite{Gupta1950,Gupta1977,ItzyksonZuber},
and demonstrate the UV divergences of the vacuum stress tensor.
Finally, we shall perform the adiabatic regularization
on the vacuum stress tensor,
and show that the resulting regularized  vacuum stress tensor is zero,
so that all the predicted physics of the Maxwell field in de Sitter space
is the same,   with or without the GF term.

The paper is organized as follows.
Sect.2, we derive, by two methods,
 the solutions of the Maxwell field
with a general GF  term in de Sitter space.
Sect. 3. we present the covariant canonical quantization.
Sect. 4, we calculate all the three parts of the stress tensor.
Sect. 5, we perform  the adiabatic regularization of
  the vacuum stress tensor.
Sect. 6   gives the  conclusion  and  discussions.
Appendix A gives the Green's functions for the Maxwell field
in the Feynman gauge,
and demonstrates its relation to  the Green's function
 of a minimally-coupling massless scalar field.
In Appendix B,   we give  analogous calculation of
the Maxwell field with a general GF  term
in the Minkowski spacetime,
which has not been fully reported in literature.
We shall use the units ($\hbar=c=1$).

\section{ The solutions of the Maxwell equations
        with $\zeta$  in de Sitter space:}

In the free Maxwell field theory,
the longitudinal and temporal components, $\bf A_{||}$ and $A_0$,
are not real radiative dynamical degree of freedom.
A simplest treatment is to take the Coulomb (radiation) gauge,
in which the longitudinal and temporal components are set to be zero,
$\nabla \cdot {\bf A} =0 = A_0$,
and  the canonical quantization is performed
only on the transverse parts.
The treatment in the Coulomb gauge is not explicitly covariant.
To achieve the  covariant canonical quantization,
one can introduce  a GF term,
so that the canonical momenta are not identically zero,
and all the four components $A_\mu$
can be regarded as being dynamical variables
without the Lorenz condition.
Nevertheless,
the GF  term will cause $A$ and $A_0$ to mix up in
their field equations,
and the solution is nontrivial.
In this section
we shall derive the solution of $A_\mu$
and the corresponding canonical momenta
of the Maxwell field with the GF  term in de Sitter space.

The metric of a flat Robertson-Walker (RW) spacetime
is written as
\be \label{metric}
ds^2=a^2(\tau)[- d\tau^2 + \delta_{ij}   dx^idx^j],
\ee
which is conformal to the Minkowski spacetime,
with  $\tau$ being the conformal time.
The Lagrangian density  of the Maxwell field with a GF  term
 in RW spacetimes  is  \cite{AdlerLiebermanNg1977}
\bl\label{Lagrangianzeta}
{\cal L}  &
 = \sqrt{-g}\Big( -\frac14 g^{\mu\rho}g^{\nu\sigma}
    F_{\mu\nu} F_{\rho\sigma}
    -\frac{1}{2\zeta}  (A^\nu\, _{; \, \nu} )^2
    \Big) ,
\el
where $F_{\mu\nu} = A_{\mu , \, \nu}- A_{\nu , \, \mu}$,
and $\zeta$ is the gauge fixing constant.
The field equation  of $A_\mu$ is
\bl
 F^{\mu\nu}\, _{;\, \nu} + \frac{1}{\zeta} (A^\nu\, _{; \, \nu})^{;\mu}=0 .
 \label{eqFmunu}
\el
Applying the covariant four-divergence upon eq.\eqref{eqFmunu}
gives $\Box (A^\nu\, _{; \, \nu}) =0$,
where $\Box \equiv - \frac{1}{a^4} \frac{\partial}{\partial \tau}
(a^2 \frac{\partial}{\partial \tau}) + \frac{1}{a^2} \nabla^2$,
so,  $(A^\nu\, _{; \, \nu})$ satisfies
the equation of a minimally-coupling massless scalar field.
In this paper,
all the four components $A_\mu$ will be formally regarded as
dynamical field variables,
and the Lorenz condition
will not be imposed as a condition on the field operators.
The equation \eqref{eqFmunu}  is written as
\bl \label{equaAm0zeta}
   \eta^{\sigma \rho} \partial_\sigma  \partial_\rho A_\mu &
      +\Big(  \frac{1}{\zeta} -1 \Big)
      \partial_\mu (\eta^{\rho\sigma} \partial_\sigma  A_\rho)
        \nn \\
& + \frac{1}{\zeta}
  \Big[ \delta_{\mu 0}
  \Big( - D \eta^{\rho\sigma } \partial_\sigma  A_\rho  + D^2 A_0 - D' A_0 \Big)
         -  D \partial_\mu A_0  \Big]  =0 ,
\el
where    $\eta^{\mu\nu}=diag \, (-1,1,1,1)$,
$D \equiv  2  a'(\tau)/a(\tau)$.
The $i$-component $A_i$ is decomposed into
\bl \label{Aidecp}
A_i  & = B_i +\partial_i A \, ,
\el
where $\partial_i B_i =0$ and
$A$ is a scalar function and  $\partial_i A$
is the longitudinal.
The canonical momenta are defined by
\bl\label{pi0}
\pi_A^\mu & = \frac{\partial \cal L}{\partial (\partial_0 A_\mu)}
  =     \eta^{\mu \sigma}
     (\partial_0  A_\sigma - \partial_\sigma A_0 )
     - \frac{1}{\zeta}  \eta^{0 \mu}
     \Big( \eta^{\rho\sigma}  \partial_\sigma  A_\rho - D A_0 \Big) ,
\el
its $0$-component is contributed by the GF  term,
\bl   \label{pi0def1}
\pi_A^0
  &  =  \frac{1}{\zeta} \,  a^2  A^\nu\, _{;\, \nu} \,
      = \frac{1}{\zeta} \Big( -(\partial_0 + D) A_0 +\nabla^2  A \Big)
\el
and the  $i$-component is
\bl
\pi_A^i  &  =  \delta^{i j} (\partial_0  A_j - \partial_j A_0 )
   = w^i +\partial^i \pi_A \,  ,  \label{decpwi}
\el
where $w^ j  = \partial_0  B_j $ is  transverse,
and
\bl
\pi_A & = \partial_0   A  -   A_0
      \label{piac}
\el
is a scalar function and its gradient $\partial^i \pi_A$
is the longitudinal.
For convenience,
in the following of this section,
we shall work with
the Fourier $k$-modes of the fields and the canonical momenta,
for instance,
$B_i(x)=\int\frac{d^3 k}{(2\pi)^{\frac32}}
   B_{i \, k}(\tau) e^{i\bf{k}  \cdot \, \bf{x}}$,  etc.
To avoid the cumbersome notation of subindex $k$,
we also use $B_i$, $A$, $A_0$, $\pi_A$, $\pi_A^0$
to represent their  $k$-modes whenever  no confusion arises
in the following.
Then,  with $\nabla^2   = -k^2$ ,
the $k$-mode of \eqref{pi0def1} is written as
\be
\pi_A^0    =  - \frac{1}{\zeta}
  \big( (\partial_0 + D) A_0 + k^2  A \big) .
  \label{pi0def}
\ee
Eq.(\ref{equaAm0zeta}) is decomposed into
the following equations in the $k$-space,
\bl
& \partial_0^2 B_i + k^2  B_i   =0 ,
\label{Bieq1mk}
\\
&  - \partial_0^2 A -\frac{1}{\zeta} k^2 A
   + (1- \frac{1}{\zeta} )\partial_0  A_0 - \frac{1}{\zeta} D A_0
      =0  ,   \label{0aA}
      \\
& - \frac{1}{\zeta}\partial_0^2  A_0  -k^2 A_0
  +\frac{1}{\zeta} ( D^2  -D' ) A_0
  + k^2 \big( (1- \frac{1}{\zeta} )\partial_0 A + \frac{1}{\zeta} D A \big)
     =0  ,   \label{Aa0}
\el
where $B_i$, $A$, and  $A_0$ stand for their $k$-modes.
The  transverse equations  \eqref{Bieq1mk}
are separated from $A$ and $A_0$,
 unaffected by the gauge fixing parameter,
and,   each $i$-component of the $k$-mode  $B_i$
has the  positive frequency solution of the following form
\be \label{f12m0solu}
B_i(\tau)  \propto
  f_k^{(\sigma)} (\tau)   = \frac{1}{\sqrt{2k}}e^{- i k\tau}  ,
\ee
where  the solution modes $f_k^{(\sigma)} $
are the same for two transverse polarizations  $\sigma =1,2$
(see \eqref{expBi}  \eqref{polariz} for a precise expression of $B_i$.)
The equation  \eqref{Bieq1mk}
and the solution \eqref{f12m0solu}
are independent of the scale factor $a(\tau)$,
and hold for a general RW spacetime,
including de Sitter space and  the Minkowski spacetime.

Eqs.(\ref{0aA})(\ref{Aa0})
are the basic second order differential equations
of $A$ and $A_0$ for a general $\zeta$,
in which $A$ and $A_0$ are mixed up.
Even in the Feynman gauge $(\zeta=1)$,
 (\ref{0aA}) (\ref{Aa0}) become
\bl
 - \partial_0^2 A -  k^2 A   -  D A_0    &  =0 ,  \label{f0aA}
 \\
 -   \partial_0^2  A_0  -k^2 A_0
  +  ( D^2  -D' ) A_0
  + k^2    D A     &  =0  ,   \label{fAa0}
\el
where $A$ and $A_0$ are still mixed up.
(When $D=0$,  eqs.\eqref{f0aA}\eqref{fAa0}
reduce to eqs.\eqref{feynmanminkwA} \eqref{feynman4thAeq}
in the Minkowski spacetime that is most discussed in literature,
 and $A$ and $A_0$ are separate.)

We shall solve eqs.(\ref{0aA}) (\ref{Aa0})
 with  a general $\zeta$ in the following.
By differentiations and algebraic combinations of (\ref{0aA}) (\ref{Aa0}),
we get  two 4th-order differential equations
\bl  \label{Aeom}
& \Big[ (1- \frac{1}{\zeta} )\partial_0
- \frac{1}{\zeta} D  \Big]
\Big( \frac{\Big( (\zeta - 1) \partial_0^3
  +   D \partial_0^2
 + k^2 (\zeta - 1) ( 2 - \zeta )\partial_0
   +  (2 - \zeta)  k^2 D \Big)  A}
   {(\zeta - 2)  D^2 -(\zeta - 1) D' - (\zeta - 1)^2 k^2} \Big)
   \nn \\
&  - \Big(  \partial_0^2  + \frac{1}{\zeta} k^2 \Big)  A  = 0  ,
\el
and
\bl  \label{A0eom}
&     \Big[ (1- \frac{1}{\zeta} )\partial_0 + \frac{1}{\zeta} D \Big]
     \Bigg( \Big[(\zeta -1) D' - D^2
     - (\zeta -1)^2 \frac{1}{\zeta}  k^2 \Big]^{-1}
     \Big[(\zeta -1) \partial_0^3 A_0 - D  \partial_0^2  A_0
   \nn \\
&   + (\zeta -1) ( D'  - D^2 - \frac{1-2\zeta }{\zeta} k^2  )A_0'
 + (\zeta -1)  ( D'' - 2 D D' ) A_0
                  \nn \\
&   + D ( D^2  -D' ) A_0
    + \frac{1-2\zeta }{\zeta} k^2 D A_0 \Big] \Bigg)
    \nn \\
&  - \frac{1}{\zeta}\partial_0^2  A_0 -k^2 A_0
  +\frac{1}{\zeta}  ( D^2  -D'  ) A_0  =0 ,
\el
which are separate for $A$ and $A_0$,
and valid for $\zeta \ne 1$.
(When  $D=0$,  eqs.\eqref{Aeom} \eqref{A0eom}
reduce to eqs.\eqref{minkwA} \eqref{4thAeq}  in the Minkowski spacetime.)

In this paper we consider de Sitter space,
  the scale factor is
\bl \label{deSittera}
a(\tau)= -\frac{1}{H\tau}, \,\,\,\,-\infty<\tau\leq \tau_1,
\el
where $H$ is a constant,
$\tau_1$ is the ending time of  de Sitter inflation, $D    =-2/\tau$.
Dropping an overall factor $\propto(1-\zeta)^2$,
eqs.\eqref{Aeom}   \eqref{A0eom}  become
\bl
&[(\zeta-1)^2k^2\tau^2-2(\zeta-3)]\tau^2A^{(4)}(\tau)-4(\zeta-3)\tau A^{(3)}(\tau)\nn
\\
&~~~+2[(\zeta-1)^2k^4\tau^4-(\zeta+1)(\zeta-3)k^2\tau^2+2(\zeta-3)]A''(\tau)\nn
\\
&~~~+4(\zeta-2)(\zeta-3)k^2\tau A'(\tau)\nn
\\
&~~~
+[(\zeta-1)^2k^4\tau^4-2\zeta(\zeta-3)k^2\tau^2-4(\zeta-3)]k^2A(\tau) =0
\label{A4ordereqSimpl0}
\el
and
\bl
&[(\zeta-1)^2k^2\tau^2-2(\zeta-3)\zeta]\tau^4A_0^{(4)}(\tau)-4(\zeta-3)\zeta\tau^3 A_0^{(3)}(\tau)\nn
\\
&~~~+[2(\zeta-1)^2k^4\tau^4+2(-3\zeta^2+6\zeta+1)k^2\tau^2+4(\zeta-3)\zeta]\tau^2 A_0''(\tau)\nn
\\
&~~~+4[(3\zeta-1)k^2\tau^2-2(\zeta-3)\zeta]\tau A_0'(\tau)\nn
\\
&~~~+[(\zeta-1)^2k^6\tau^6
+2(-2\zeta^2+3\zeta+1)k^4\tau^4
+4(-2\zeta^2+3\zeta+1)k^2\tau^2+8\zeta(\zeta-3)]A_0(\tau) =0
\label{A04ordereqSimpl0}
\el
where
$A^{(4)}(\tau) \equiv \partial^4 A/\partial\tau^4$,
$A^{(3)}(\tau) \equiv \partial^3 A/\partial\tau^3$, etc.
These are 4th-order differential equations of $A_0$ and $A$,
  valid for a general $\zeta$.
Setting $\zeta=1$, they  reduce to
the 4th-order differential  equations in the Feynman gauge
\bl
&  \tau^2 A^{(4)}  + 2 \tau A^{(3)}
 +2(k^2 \tau^2-1) A''  +2 k^2 \tau A'
\nn \\
& ~~~~~~ +k^2 (k^2 \tau^2  + 2 ) A =0 ,
  \label{FymeqA}
\\
& \tau^4 A_0^{(4)}  +2 \tau^3 A_0^{(3)}
+ 2 \tau^2  ( k^2 \tau^2-  1  ) A_0''
+2 (k^2 \tau^2+2 ) \tau A_0 '
\nn \\
& ~~~~~~   + (k^4 \tau^4 +2 k^2 \tau^2 -4 ) A_0   =0 .
 \label{FymeqA0}
\el
The positive frequency solutions of
eqs.\eqref{A4ordereqSimpl0}  \eqref{A04ordereqSimpl0}
for a general $\zeta $ are obtained
\bl
A & = b_1  \frac{1}{a(\tau)} \frac{i}{k}
   (1-\frac{i}{k\tau }) \frac{1}{\sqrt{2k}} e^{-i k\tau }
   - b_2 \frac{ ( {3 }- \zeta  ) (k \tau +i)
   e^{2 i k \tau } \text{Ei}(-2 i k \tau ) -{3 i } +2 i \zeta  }
    {3 k}
    \frac{1}{\sqrt{2 k}} e^{-i k \tau } , \label{H12}
   \\
 A_{0} & =   b_1 \frac{1}{a(\tau)} \frac{1}{\sqrt{2 k}} e^{-i k \tau}
    - b_2 \frac{ \left(3 i  -i \zeta\right)
    k^{2} \tau^{2}
      e^{2 i k \tau} \operatorname{Ei}(-2 i k \tau)
     +\zeta(k \tau-i) }{3 k \tau} \frac{1}{\sqrt{2 k}} e^{-i k \tau},
     \label{A0bar}
\el
where $\text{Ei}(z)\equiv -\int_{-z}^{\infty }  t^{-1} e^{-t} \, dt$
is the exponential-integral function,
and  the coefficients $b_1,b_2$ are dimensionless complex constants.
(Refs.\cite{JimenezMaroto2009,JimenezMaroto2010} gave a solution
which seems to correspond
to the special case $\zeta=1$  of our \eqref{H12} \eqref{A0bar}.)
We have chosen the same set of coefficients
$(b_1, b_2)$ for $A$ and $A_0$
so that they satisfy
the basic 2nd-order equations  \eqref{0aA} \eqref{Aa0}.
At the classical level, $(b_1,b_2)$ are arbitrary.
The $b_1$ part will be referred to as the homogeneous solution,
and the $b_2$ part as the inhomogeneous solution,
and the terminologies
``homogeneous" and  ``inhomogeneous"
will be clear later around \eqref{Abareq} $\sim$ \eqref{barA0sol}.
The complex conjugates of
\eqref{H12} \eqref{A0bar}
are the independent, negative frequency solutions.
Although $A$ and $A_0$ respectively
have four solutions (the Wronskians being nonzero),
but  $A$ and $A_0$ in \eqref{H12}  \eqref{A0bar}
share the same set $(b_1,b_2)$.
We have checked that
the  respective homogeneous and inhomogeneous parts
in \eqref{H12}  \eqref{A0bar}
satisfy the basic 2nd-order equations \eqref{0aA} \eqref{Aa0},
as well as  the 4th-order equations
\eqref{A4ordereqSimpl0}  \eqref{A04ordereqSimpl0}.
When setting  $\zeta=1$, \eqref{H12}  \eqref{A0bar}
 reduce to the solutions of \eqref{FymeqA} \eqref{FymeqA0} in the Feynman gauge.
Plugging   \eqref{H12}  \eqref{A0bar} into
the definitions \eqref{piac} \eqref{pi0def}
gives the canonical momenta
\bl
\pi_A &  = - b_2 \frac{i}{k \tau }  \frac{1}{ \sqrt{2 k }} e^{-i k \tau }
 =   b_2  \frac{i H}{k} a(\tau) \frac{1}{ \sqrt{2 k }} e^{-i k \tau } ,
  \label{pides}
              \\
\pi^0_A  & = b_2 \frac{ i -k \tau}{k \tau ^2} \frac{1}{\sqrt{2k}} e^{-i k \tau }
=     b_2 H  a(\tau)  (1 -\frac{i}{k \tau})  \frac{1}{\sqrt{2k }} e^{-i k \tau }
    , \label{pi0des}
\el
which are contributed only by the inhomogeneous part of $A$ and $A_0$,
and are independent of $\zeta$.
It should be remarked  that  the  positive frequency
 ($\propto e^{-i k \tau }$) modes
\eqref{H12} \eqref{A0bar} \eqref{pides} \eqref{pi0des}
will not evolve into the negative frequency modes ($\propto e^{i k \tau }$)
during the de Sitter expansion
prescribed by  \eqref{metric} \eqref{deSittera}.
Note that the dimension $[A_0]=k[A]$ and $ [\pi^0_A] = k [\pi _A]$.

The solutions  \eqref{H12} \eqref{A0bar} in the de Sitter space
will reduce   to  the solutions  in the Minkowski spacetime.
But,  if one naively took $a=1$ and high $k$
in \eqref{H12} \eqref{A0bar},
one would come up with an incorrect  claim that
the Minkowski limit can be obtained at only for $\zeta = -3$.
In fact, $a(\tau)$ and its time derivatives
are implicit in  \eqref{H12} \eqref{A0bar}.
An appropriate procedure of taking limit to the Minkowski spacetime
is:
Setting $D=0$ in eqs.\eqref{Aeom} \eqref{A0eom}
leads to eqs.\eqref{minkwA} \eqref{4thAeq}
in the  Minkowski spacetime,
and the solutions are listed in Appendix B.

The solutions \eqref{H12} \eqref{A0bar} \eqref{pides} \eqref{pi0des}
can also be derived in another way as the following.
First, by applying $\partial_0$ and combinations on
the basic  equations (\ref{0aA}) (\ref{Aa0}),
we arrive at the equations  of  $\pi_A$ and $\pi^0_A$,
\bl
(\partial_0^2   - D \partial_0   +k^2 ) \pi_A & =0 ,  \label{piAeq1}    \\
( \partial_0^2  - D \partial_0 - D'  +k^2 ) \pi^0_A  &  =0 , \label{pi0Aeq}
\el
which are independent of $\zeta$.
By rescaling  $\pi_A  =   a \bar \pi_A $,
$\pi_A^{0}  =   a \bar \pi_A^{0}$,
eqs.\eqref{piAeq1} \eqref{pi0Aeq}  become
\bl \label{eqY3m0desitter}
\bar \pi_A \, '' +  k^2  \bar \pi_A  & =0 ,
\\
\bar \pi_A^{0} \, ''     + (k^2 - \frac{2}{\tau^2} ) \bar \pi_A^{0} &  =0 ,
\label{barpi0eq}
\el
and the normalized solutions  are
\bl \label{Y3m0}
\bar \pi_A   & = b_2 H \frac{i}{k} \frac{1}{\sqrt{2k}}e^{- i k\tau} ,
\\
\label{Y0sol}
\bar \pi_A^{0} & = b_2 H \frac{1}{\sqrt{2k}} (1- \frac{i}{k \tau})e^{- i k\tau} .
\el
Multiplying the above by $a(\tau)$ gives
the solutions \eqref{pides}  \eqref{pi0des}.
Note that eq.\eqref{eqY3m0desitter} of $\bar \pi_A$
is the same as the equation of
a rescaled conformally-coupling massless scalar field,
and eq.\eqref{barpi0eq} of $\bar \pi^0_A$ is
the same as the equation of
a rescaled minimally-coupling massless
scalar field \cite{ZhangWangYe2020}.
Next,
applying  $\partial_0 $ on the definitions (\ref{piac}) (\ref{pi0def})
and by combinations, we arrive at
\bl
 \partial_0^2 A  + D \partial_0 A   +k^2 A
  &  = (\partial_0 + D)  \pi_A  - \zeta \pi_A^0  , \label{AEu2}
  \\
 \partial_0^2  A_0 +   D A_0'  + D' A_0   + k^2   A_0
     &  = -( k^2  \pi_A + \zeta \partial_0   \pi_A^0) , \label{A0epi}
\el
which are  the second order differential equations of $A$ and $A_0$
 with the nonhomogeneous term as the source.
(The homogeneous equations of \eqref{AEu2}  \eqref{A0epi}
are just the equations of $A$ and $A_0$ of
Maxwell theory without the GF term
under the Lorenz condition $ A^{ \mu} \, _{;\mu} =0$.)
By  rescaling $A_0 = \frac{1}{a} \bar A_0$
and $A = \frac{1}{a} \bar A$,
eqs.(\ref{A0epi}) \eqref{AEu2} become
\bl
\bar A ''  + (k^2   -  \frac{2}{\tau^2} ) \bar A
    &  =\Pi(\tau) ,
    \label{Abareq}
      \\
\bar A_0 ''  +  k^2      \bar A_0
   & =  \Pi_{0}(\tau) ,
   \label{A0bareq}
\el
where the nonhomogeneous terms  are
\bl
\Pi(\tau)
& \equiv    a \big((\partial_0 + D)  \pi_A  - \zeta \pi_A^0 \big) ,
 \label{solpiA}
 \\
\Pi_{0}(\tau) & \equiv
  -a \big( k^2  \pi_A + \zeta \partial_0 \pi_A^0 \big)  ,
  \label{Pia0def}
\el
which are known from the given $\pi_A$ and $\pi_A^0$.
The homogeneous solutions of \eqref{Abareq} \eqref{A0bareq}
 are simply given by
\bl
\bar  A_{h}(\tau) &  =  \frac{i}{k} \frac{1}{\sqrt{2k} }
 (1-\frac{i}{k\tau }) e^{-i k\tau } , \label{barA1}
 \\
\bar  A_{0h}(\tau) &  = \frac{1}{\sqrt{2k}} \,  e^{-ik\tau} ,
\label{barA01}
\el
which correspond to the $b_1$ part of the solutions \eqref{H12}  \eqref{A0bar},
and  the   Wronskians are
\bl
W [\tau] & = \bar A_{h} \bar A^{*'}_{h}  - \bar A_{h}' \bar A^*_{h}
 =\frac{i}{k^2},
 \\
W_0[\tau] & = \bar A_{0h} \bar A^{*'}_{0h}  - \bar A_{0h}' \bar A^*_{0h} =i .
\el
Interestingly,  the homogeneous equation \eqref{Abareq}
and the solution \eqref{barA1} of  $\bar A_h$
are similar to \eqref{barpi0eq}   \eqref{Y0sol} of $\bar \pi^0_A$,
and,
the homogeneous equation \eqref{A0bareq}
and the solution \eqref{barA01} of $\bar A_{0h}$
are similar to \eqref{eqY3m0desitter}  \eqref{Y3m0}
 of $\bar \pi_A$ \cite{ZhangWangYe2020}.
By the standard formulae of the inhomogeneous equations,
we obtain the inhomogeneous
solution of (\ref{Abareq}) (\ref{A0bareq})
\bl \label{barAsol}
 \bar A (\tau) & = -  \bar  A_h(\tau) \int^\tau  d\tau'
       \frac{\Pi(\tau') \bar  A^*_h (\tau')}{W}
     + \bar  A_h^* (\tau) \int^\tau   d\tau'
        \frac{\Pi(\tau') \bar  A_h (\tau')}{W},
         \nn \\
& =   b_2  \frac{\left(( 3 - \zeta ) (k \tau +i)
   e^{2 i k \tau }  \text{Ei}(-2 i k \tau ) - 3 i   +2 i \zeta   \right)}
   {3  H k  \tau }  \frac{1}{\sqrt{2 k}} e^{-i k \tau } ,
\\
 \bar A_{0} (\tau) & = \bar  A_{0h}(\tau) \int^\tau  d\tau'
    \frac{ - \Pi_{0}(\tau') \bar  A_{0h}^* (\tau')}{W_0}
   + \bar  A_{0h}^*(\tau) \int d\tau'
      \frac{ \Pi_{0}(\tau') \bar  A_{0h}(\tau')}{W_0}
      \nn \\
& = b_2 \frac{\left((3 i  -i \zeta ) k^2 \tau ^2
  e^{2 i k \tau } \text{Ei}(-2 i k \tau )
  +\zeta  (k \tau -i)\right)}{3  H k  \tau ^2}
   \frac{1}{\sqrt{2k}} e^{-i k \tau } . \label{barA0sol}
\el
After rescaling by $1/a(\tau)$,
the  sum of  \eqref{barA1} and \eqref{barAsol}
recovers the solution $A$ in \eqref{H12},
and the sum of \eqref{barA01}  and  \eqref{barA0sol}
recovers  the solution of $A_0$ in  \eqref{A0bar}.
As we shall see later,
the  complicated,  inhomogeneous parts of $A$ and $A_0$
will simply cancel in the expectation value of the stress tensor.

We analyze the gauge transformations of the Maxwell field,
and examine the consequential changes  on the solutions.
The Maxwell field without the GF term
is invariant under the gauge transformation
$A_\mu \rightarrow A_\mu' \equiv  A_\mu +\theta_{,\mu}$
with $\theta$ being an arbitrary scalar function,
each component transforms as
\bl
B_i \rightarrow B_i' & =  B_i \, ,
\nn \\
A  \rightarrow A ' &  \equiv  A  +\theta \, ,
\nn \\
A_0 \rightarrow A_0' & \equiv  A_0 +\theta_{,\, 0} \, .
\nn
\el
When the GF term  $\propto (\nabla^\mu A_\mu )^2$ is present,
the Lagrangian   \eqref{Lagrangianzeta}
and the field equation \eqref{eqFmunu}
are invariant only under a residual gauge transformation
with   $\theta$  satisfying the following  equation
\bl \label{thetaequ}
\Box \theta \equiv \nabla^\nu\nabla_\nu \theta=0 .
\el
This is also the equation of
a minimally coupling massless scalar field \cite{ZhangWangYe2020},
and its  $k$-mode  equation is
\be \label{thetaeq}
  \theta_k '' + D  \theta_k '  + k^2 \theta_k =0 .
\ee
In de Sitter space the $k$-mode solution is
\bl
\theta_k(\tau) = C \frac{1}{a(\tau)} \frac{i}{k} (1-\frac{i}{k\tau})
           \frac{1}{\sqrt{2k}} e^{-ik\tau} ,
           \label{soltheta}
\el
with $C$ being an arbitrary complex constant.
The function  $\theta_k$ in \eqref{soltheta}
is of the same form as the homogeneous solution $ A_h$ of \eqref{H12},
and its  time derivative is
\be
 \theta_{k, \, 0}   = C \frac{1}{a(\tau)}
  \frac{1}{\sqrt{2k}}  e^{-ik \tau} ,
\ee
whose form is the same as the homogeneous solution $A_{0h}$ of \eqref{A0bar}.
Thus, under the residual gauge transformation,
 the longitudinal and temporal $k$-modes transform as
\bl
A_{k}   &  \rightarrow
     A_{ k} + C \frac{1}{a(\tau)}  \frac{i}{k}
       (1-\frac{i}{k\tau}) \frac{1}{\sqrt{2k}} e^{-ik\tau} ,
       \label{Atrsf}
  \\
A_{0\, k} & \rightarrow
        A_{0\, k} +  C \frac{1}{a(\tau)}
       \frac{1}{\sqrt{2k}}  e^{-ik\tau} .
          \label{A0trsf}
\el
Comparing with the solutions \eqref{H12} \eqref{A0bar}
of $A$ and $A_0$,
the residual gauge transformation
\eqref{Atrsf} \eqref{A0trsf}
amounts to a change of the homogeneous parts of $A$ and $A_0$
as the following
\bl
b_1 \rightarrow b_1 ' = b_1 + C.
\el
Under the residual gauge transformation
the canonical momenta are invariant
\bl
\pi_A   \rightarrow \,  &  \partial_0  ( A +\theta)
 -  ( A_0 +\theta_{,0})
  =       \pi_A ,   \label{pitrs}
\\
\pi^0_A  \rightarrow  &
  - \frac{1}{\zeta} \Big(  \partial_0  ( A_0+\theta_{,\,0})
    + D ( A_0+\theta_{,\,0}) + k^2 ( A+\theta) \Big)
    = \pi^0_A .  \label{pi0trsf}
\el
This invariant  property is consistent with the fact that
the solutions $\pi_A $ and $\pi^0_A $ in \eqref{pides}  \eqref{pi0des}
are contributed  only by the inhomogeneous parts of $A$ and $A_0$,
and, therefore,  unaffected by any change of the homogeneous parts.

As  we shall show in the next section,
a consistent covariant canonical quantization
requires that the homogeneous part of $A$ and $A_0$ be nonvanishing,
$b_1 \ne 0$, $b_1'\ne 0$.
Therefore,  at the quantum level,
the parameter $C$ of residual gauge transformation
will be further restricted.

\section{ \bf The covariant canonical quantization
of Maxwell field with general $\zeta$ in de Sitter space  }

After obtaining all  the  $k$-modes
\eqref{f12m0solu} \eqref{H12} $\sim$ \eqref{pi0des}
for general $\zeta$ in de Sitter space,
we shall implement the covariant canonical quantization.
This procedure will constrain the coefficients for each mode,
and restrict the residual gauge transformation as well.
The field operators are required to satisfy
the  equal-time covariant canonical commutation relations,
\bl
& [A_\mu(\tau, {\bf x}),\pi_A^{\nu}(\tau, {\bf x'})]
           =i g^{\nu}_{~\mu} \delta({\bf x-x'}),
  \label{N36}
\el
with $g^{\nu}_{~\mu} =\delta^{\nu}_{~\mu}$,
the other commutators  vanish.
The $ij$-component of commutation relations
can be decomposed into
\bl \label{AiPiA}
[A_i, \pi_A^j  ]
& = [(B_i+A_{,i}), (w^j + \pi_{A}^{\, , j}) ] \nn \\
   &  =[B_i,  w^j] + [\partial_i A, \partial^j \pi_A ] ,
\el
where the transverse and longitudinal
are independent,  and commute with each other.

The transverse components $B_i$ in de Sitter space
are simply the same as in the Minkowski spacetime.
We write the operators of
the  transverse fields and canonical momenta as
\bl \label{expBi}
B_i   ({\bf x},\tau)
& =  \int\frac{d^3k}{(2\pi)^{3/2}}
  \sum_{\sigma=1}^2 \epsilon^{\sigma}_{i}(k)
  \left[ a^{( \sigma)} _{\bf k}  f_k^{(\sigma)}(\tau)   e^{i\bf{k} \cdot\bf{x}}
   +a^{( \sigma)\dagger}_{\bf k} f_k^{(\sigma)*}(\tau)
   e^{-i\bf{k} \cdot\bf{x}}\right] ,
\\
w^i (\tau, {\bf x})
&  =  \int\frac{d^3k}{(2\pi)^{3/2}}
  \sum_{\sigma=1}^2 \epsilon^{\sigma}_{i}(k)
  \left[ a^{( \sigma)} _{\bf k}  f_k^{(\sigma)'}(\tau) e^{i\bf{k} \cdot\bf{x}}
   +a^{( \sigma)\dagger}_{\bf k} f_k^{(\sigma)*'}(\tau)
   e^{-i\bf{k} \cdot\bf{x}}\right],
   \label{wjexp}
\el
where the modes $f^{(1,2)}_k(\tau)$ are given by (\ref{f12m0solu}),
the commutators of the transverse creation
 and annihilation operators are
\bl \label{aa12com}
& [a_{\bf k}^{(\sigma)}, a_{\bf k'}^{(\sigma')\dagger}]
=\eta^{\sigma\sigma '} \delta^{(3)}({\bf k-k'}),
    ~~~~~~~~~ (\sigma=1,2)  ,
\el
the transverse  polarizations satisfy
\be
\sum_{i=1,2,3}k^i \epsilon^\sigma_i(k) =0, ~~~
\sum_i \epsilon^\sigma_i(k) \epsilon^{\sigma'}_i(k)=\delta^{\sigma\sigma '},
~~~
\sum_{\sigma=1,2}   \epsilon^\sigma_i(k) \epsilon^{\sigma}_j(k)
   = \delta_{ij}-\frac{k_i k_j}{k^2} \, .
   \label{polariz}
\ee
Calculation yields
\bl
[B_i (\tau, {\bf x}), w^j (\tau, {\bf x'})]
=  &    i \delta_{i j} \delta^{(3)}({\bf x-x'})
  -  i  \int \frac{d^3k  }{(2\pi)^{3}} \Big( \frac{k_i k_j}{k^2} \Big)
    e^{i \bf k\cdot x} e^{-i \bf k\cdot x'} . \label{BiWico}
\el

The longitudinal $A$  and temporal  $A_0$
are mixed up in the basic equations  (\ref{0aA}) (\ref{Aa0})
  in de Sitter space,
so their field operator expansions
are written  as the following
\bl
A = & \int\frac{d^3k}{(2\pi)^{3/2}}
  \Big[ (a^{(0)}_{\bf{k}}A_{1k}+a^{(3)}_{\bf{k}}A_{2k})
   e^{i\bf{k}\cdot \bf{x}}+h.c.\Big] ,\label{N26}
   \\
A_0  = & \int\frac{d^3k}{(2\pi)^{3/2}}
 \Big[ (a^{(0)}_{\bf{k}}A_{01k}+a^{(3)}_{\bf{k}}A_{02k})
    e^{i\bf{k}\cdot \bf{x}}+h.c.\Big] ,  \label{N25}
\el
where  $a_{\bf{k}}^{(3)}$ and $a_{\bf{k}}^{(0)}$
are the annihilation operator
of the respective longitudinal and temporal field,
and satisfy
\bl \label{commaa}
[a^{(0)}_{\bf{k}},a^{(0)\dag}_{\bf{k}'}]
& =\eta^{00} \delta^{(3)} ({\bf k-k'}) = - \delta^{(3)} (\bf k-k'),
   \\
[a^{(3)}_{\bf{k}},a^{(3)\dag}_{\bf{k}'}]
   & =\eta^{33}\delta^{(3)} ({\bf k-k'})  = \delta^{(3)} ({\bf k-k'}) .
    \label{commaa00}
\el
 \eqref{aa12com}  \eqref{commaa} \eqref{commaa00}
 together constitute  the covariant commutator
\bl
[a^{(\mu)}_{\bf{k}},a^{(\nu)\dag}_{\bf{k}'}]
& =\eta^{\mu\nu} \delta^{(3)} (\bf k-k') ,
 \label{commaamunu}
\el
which  is independent of the gauge parameter $\zeta$.
The  longitudinal  and temporal  $k$-modes in \eqref{N25} \eqref{N26}
are chosen to be
\bl
A_{1k}
= & c_1\frac{1}{a(\tau)} \frac{i}{k}
   \frac{1}{\sqrt{2k}}(1-\frac{i}{k\tau }) e^{-i k\tau }
      -c_2\frac{\left(( {3 }- \zeta  ) (k \tau +i)
   e^{2 i k \tau } \text{Ei}(-2 i k \tau ) -{3 i } +2 i \zeta \right)} {3 k}
    \frac{1}{\sqrt{2 k}} e^{-i k \tau } , \label{N28}
\\
A_{2k}
= & m_1\frac{1}{a(\tau)} \frac{i}{k}
   \frac{1}{\sqrt{2k}}(1-\frac{i}{k\tau}) e^{-i k\tau }
   -m_2\frac{\left(({3 }- \zeta )(k \tau +i) e^{2 i k \tau}
    \text{Ei}(-2 i k \tau) -{3i}+2 i \zeta \right)}
    {3 k}
    \frac{1}{\sqrt{2 k}} e^{-i k \tau } , \label{N30}
\el
\bl
A_{01k}
=  & c_1 \frac{1}{a(\tau)} \frac{1}{\sqrt{2k}}  e^{-ik\tau}
    - c_2\frac{\left((3 {i }-i \zeta ) k^2 \tau ^2
    e^{2 i k \tau } \text{Ei}(-2 i k \tau )
      +\zeta (k \tau -i)\right)}{3 k \tau}
      \frac{1}{\sqrt{2k}} e^{-i k \tau} , \label{N27}
\\
A_{02k}
= & m_1 \frac{1}{a(\tau)} \frac{1}{\sqrt{2k}}  e^{-ik\tau}
    - m_2\frac{\left((3 {i }-i \zeta ) k^2 \tau ^2  e^{2 i k \tau }
    \text{Ei}(-2 i k \tau )
      +\zeta (k \tau -i)\right)}{3 k \tau} \frac{1}{\sqrt{2k}}
       e^{-i k \tau}  , \label{N29}
\el
where  $c_1, c_2, m_1,  m_2$ are dimensionless complex coefficients,
and will be subject to some constraints by the canonical quantization.
From the expansions  \eqref{N26} \eqref{N25}
together with \eqref{N28} $\sim$ \eqref{N29}
follow the expansions  of
the canonical momentum operators
\bl
\pi_A
&=\int\frac{d^3k}{(2\pi)^{\frac32}}
  \Big( \big( a_{\bf{k}}^{(0)} \pi_{A1k}
  +a_{\bf{k}}^{(3)}\pi_{A2k} \big)
         e^{i\bf{k}\cdot\bf{x}} + h.c.\Big) \label{N33} ,
         \\
\pi_A^0
&=\int\frac{d^3k}{(2\pi)^{\frac32}}
\Big(\big(a_{\bf{k}}^{(0)} \pi_{A1k}^{0}
  +a_{\bf{k}}^{(3)}\pi_{A2k}^{0}\big)
          e^{i\bf{k}\cdot\bf{x}}+h.c.\Big) ,
   \label{N31}
\el
where the $k$-modes of the longitudinal and temporal canonical momenta
are found to be
\bl
\pi_{A1k} &    = c_2\frac{-i }{ k  \tau } \frac{1}{\sqrt{2k} } e^{-i k \tau }  ,
     \label{piexc} \\
\pi_{A2k} &    = m_2\frac{-i }{ k  \tau } \frac{1}{\sqrt{2k} } e^{-i k \tau } ,
                 \label{N34}
                \\
\pi_{A1k}^{0} &  =  c_2 k \Big( - \frac{1}{k \tau} + \frac{i}{ k^2 \tau ^2} \Big)
                   \frac{1}{\sqrt{2k}} e^{-i k \tau } ,
                   \label{piA1slo}
                   \\
\pi_{A2k}^{0} &  =  m_2 k \Big( - \frac{1}{k \tau} + \frac{i}{ k^2 \tau ^2} \Big)
                   \frac{1}{\sqrt{2k}} e^{-i k \tau } .
    \label{N32slo}
\el
These canonical momentum $k$-modes  are  contributed by only
the inhomogeneous part of \eqref{N28} $\sim$ \eqref{N29}.
There are relations among the modes
\bl
m_2\pi_{A1k}=c_2\pi_{A2k} ,
     \label{relcona}
     \\
m_2\pi_{A1k}^{0} = c_2\pi_{A2k}^{0}, \label{relcon}
\el
which will be used  in Sect 4
to simplify the calculation of the stress tensor.
$\pi_A \ne 0$ and $\pi_A^0 \ne 0$,
require that  $c_2\ne 0$ and $ m_2 \ne 0$.

Substituting the operators
\eqref{N26} \eqref{N25}
 \eqref{N33} \eqref{N31}
into each component of  \eqref{N36},
and using the commutator \eqref{commaamunu},
by lengthy calculation,
we obtain the following constraints upon the coefficients
\bl
 &  |m_1|^2 -|c_1|^2 =0 , ~~~~~~ ~~~ \text{(from  $[A_0,A_i]$)}
 \label{desitterc1m1}
 \\
  \label{N44}
&  |m_2|^2-|c_2|^2=0, ~~~~~~~~~ \text{(from $ [A_0,\pi_A^{0}]$)}
\\
& m_2m_1^*-c_2c_1^*=-i k/H  , ~~~ \text{(from $ [A_0,\pi_A^{0}]$)}
\label{N442}
\el
other commutators give no new constraint.
It is seen that
$c_1\ne 0$, $m_1 \ne 0$, $c_2\ne 0$, $m_2 \ne 0$.
This means that
 both the homogeneous and inhomogeneous parts
of the modes  \eqref{N28} $\sim$ \eqref{N29}
must be present in order to
achieve  the covariant canonical quantization \eqref{N36}.
There are many choices to satisfy the set of constraints
\eqref{desitterc1m1} \eqref{N44}  \eqref{N442}.
For instance, we take  the following specific values,
\be
 c_1=m_1=1, ~~~~ c_2= i\frac{k}{2H},
  ~~~~  m_2  =  - i\frac{k}{2H} , \label{allowc12}
\ee
which will be consistent with
those   in the Minkowski spacetime.

Another implication of  the constraints
\eqref{desitterc1m1} \eqref{N44}  \eqref{N442}
is that,
in order to ensure the nonvanishing homogeneous part of $A$ and $A_0$,
the residual gauge transformation will be  further restricted.
Under   the residual gauge transformation
\eqref{Atrsf} \eqref{A0trsf},
the  $k$-modes $(A_{1k}, A_{01k})$ and $(A_{2k},A_{02k})$
change as
\bl
 A_{1k} \rightarrow  \tilde{A}_{1k}
 & =A_{1k} +C\frac{1}{a}\frac{i}{k}(1-\frac{i}{k\tau})
     \frac{1}{\sqrt{2k}}e^{-i k\tau},\label{X266}
\\
 A_{01k}\rightarrow  \tilde{A}_{01k}
& =A_{01k}+C\frac{1}{a}\frac{1}{\sqrt{2k}}e^{-i k\tau},\label{X277}
\\
 A_{2k}\rightarrow  \tilde{A}_{2k}
& =A_{2k}+M\frac{1}{a}\frac{i}{k}(1-\frac{i}{k\tau})
   \frac{1}{\sqrt{2k}} e^{-i k\tau}  ,  \label{X288}
\\
 A_{02k}\rightarrow  \tilde{A}_{02k}
& =A_{02k}+M\frac{1}{a}\frac{1}{\sqrt{2k}}e^{-i k\tau} ,
  \label{X299}
\el
where   $C$ and $M$ are two constants
and shift only the coefficients
of the homogeneous parts
\bl
 c_1 \rightarrow c'_1 & = c_1 +C,
 \\
  m_1 \rightarrow m'_1 & = m_1 +M .
\el
In analog to the constraints
\eqref{desitterc1m1}  $\sim$  \eqref{N442},
the new coefficients also obey the following  constraints
\bl
&|m'_1|^2-|c'_1|^2=0,\label{Newsecondconstrain}
\\
&|m'_2|^2-|c'_2|^2=0,\label{Newfistconstrain}
\\
&m'_2 m_1^{'*}-c'_2 c_1^{'*}=-i\frac{k}{H} ,  \label{Newthirdconstrain}
\el
which leads to the following restriction on the constants $C$ and $M$:
\bl
&  |M|^2-|C|^2+ 2Re(m_1^*M-c_1^*C)   =0,
 \label{cetrans}
\\
&  m_2M^*-c_2C^*  =0.\label{forgaugetrans}
\el
For the   choice \eqref{allowc12},
the restriction \eqref{cetrans}  \eqref{forgaugetrans} becomes
\be \label{reggtf}
C    = - M = i \,  r,
\ee
where $r$ is an arbitrary  real number.
As a result,
the  homogeneous parts will not be transformed  to zero
\bl
c'_1     = 1 + i r \ne 0,
~~~~
m'_1    = 1 -  i r  \ne 0.
\label{quantumtraf}
\el
We call the residual gauge transformation
with the restriction \eqref{cetrans} \eqref{forgaugetrans},
or \eqref{reggtf},
the quantum residual gauge transformation.
It is required by the covariant canonical quantization,
and is only a subset of the residual gauge transformation
\eqref{Atrsf} \eqref{A0trsf} at the classical level.

\section{ The stress tensor of the Maxwell field
       with the gauge fixing term  in de Sitter space }

The stress tensor serves as the source of the  Einstein equation.
Given the action $S[A^\mu]=\int {\cal L} \, d^4x$,
the stress tensor is defined by
$T_{\mu\nu} = -\frac{2}{\sqrt{-g}}\frac{ \delta S}{\delta g^{\mu\nu}}$,
which is covariant.
Variation gives the stress tensor of
the Maxwell field with the GF  term,
\bl \label{Tmunudes}
T  _{\mu\nu} & = F_{\mu\lambda} F_{\nu}\, ^{\lambda}
         -\frac14 g_{\mu\nu} F_{\sigma \lambda} F^{\sigma \lambda}  \nn \\
& +  \frac{1}{\zeta} \Big[\frac{1}{2} g_{\mu\nu} (A^\sigma\, _{;\sigma})^2
  +  g_{\mu\nu} A^\lambda  A^\sigma\, _{; \, \sigma  \lambda}
  - A^\sigma\, _{; \, \sigma  \mu}A_\nu
  - A^\sigma\, _{; \, \sigma  \nu}A_\mu  \Big] ,
\el
and the trace  of the stress tensor is
$T^\mu \,_\mu
 = \frac{2}{\zeta} ( A^\lambda A^\sigma\, _{; \, \sigma} )_{;  \lambda}$
 which is contributed by  the GF  term only.
The corresponding energy density and pressure consist of three parts:
\bl \label{rhofl}
\rho   & = - T^0\,_0 = \rho^{TR} + \rho^{LT} + \rho^{GF} ,
          \\
 p  &  = \frac13 T^j\, _j  = p^{TR} + p^{LT} + p^{GF} .
 \label{pfl}
\el
The transverse  stress tensor is
\bl\label{trstr}
\rho^{TR} & =  3 p^{TR}
= \frac{1}{2} a^{-4} \Big( B_j ' B_j' + B_{i,j} B_{i,j}  \Big) ,
\el
which has an extra factor $a^{-4}$
to that in the Minkowski  spacetime.
This part  corresponds to the Maxwell field without the gauge term
in the Coulomb gauge.
Since $B_j$ is independent of the gauge fixing constant $\zeta$
and invariant under the residual gauge transformation,
so are $\rho^{TR}$ and $p^{TR}$.
The longitudinal-temporal (LT) stress tensor is
\bl
\rho^{LT} &  = 3p^{LT}  = \frac{1}{2} a^{-4}
          \Big( A'_{,j} A'_{,j}  +A_{0,j}A_{0,j}
           -   2 A_{0,j}  A_{,j 0}  \Big)
             \nn \\
&  =\frac{1}{2} a^{-4}   \partial_i\pi_A\partial^i \pi_A   ,
  \label{longittempml}
\el
which is written in terms of the longitudinal canonical momentum  $\pi_A$.
Since $\pi_A$ is  independent of  $\zeta$
and invariant under the residual gauge transformation \eqref{pitrs},
so are  $\rho^{LT}$ and $p^{LT}$.
The GF  stress tensor is
\bl
\rho^{GF}  & =   \frac{1}{a^4} \Big[ - \frac{1}{2} \zeta  ( \pi_A^0  )^2
      - A_0 \Big(  \partial_0 \pi_A^{0}   - D \pi_A^0  \Big)
      - A_{,j} \pi^0_{A\, ,j}   \Big]  ,     \label{rhoGFml}
  \\
p^{GF}  & =  \frac{1}{a^4 }
  \Big[\frac{1}{2} \zeta (  \pi_A^0  )^2
  - A_0 \Big(  \partial_0 \pi_A^{0}   - D \pi_A^0  \Big)
  + \frac13 A_{,j} \pi^0_{A\, ,j}  \Big] .  \label{GFpreml}
\el
This part comes from variant of the GF term
$- \frac{1}{2\zeta} \sqrt{-g} (A^\nu\, _{; \, \nu})^2$
with respect to the metric $g_{\mu\nu}$,
and involves  $\pi_A^0$,  $A$ and $A_0$.
At the classical level,
$\rho^{GF}$ and $p^{GF}$ in  \eqref{rhoGFml} \eqref{GFpreml}
apparently  depend on $\zeta$.
Besides, since $A$ and $A_0$ vary  under the residual gauge transformation,
$\rho^{GF}$ and $p^{GF}$ seem to vary too.
Later we shall see that
the expectation values of the operators
$\rho^{GF}$ and $p^{GF}$ in the GB state
are independent of $\zeta$,
and invariant under  the quantum residual gauge transformation.

In the above the stress tensor  of the Maxwell field
is still a quantum operator.
To be a source of the Einstein equation,
its  expectation value in quantum states is pertinent
\cite{UtiyamaDeWitt1962,DeWitt1975,
ParkerFulling1974,FullingParkerHu1974,ZhangYeWang2020}.
We now calculate  the expectation value of
the stress tensor.
In a state $|\phi\rangle$  of the transverse field,
using the property \eqref{polariz} of transverse  polarizations,
 we obtain
the expectation value of the transverse  stress tensor
\bl \label{rhomasslessdeSitter}
\langle \phi|  \rho^{TR} |\phi \rangle
 = & 3 \langle \phi|p^{TR} |\phi \rangle = \frac{1}{2} a^{-4}
  \langle \phi |  \Big( B_j ' B_j' +  B_{i,j} B_{i,j}  \Big)|\phi  \rangle
  \nn \\
=&   \int^{\infty}_0   \rho^{TR}_k   \frac{d k}{k}
  + \int\frac{d k}{k}   \rho^{TR}_k
   \sum_{\sigma=1,2}
  \langle \phi|a_{\bf k}^{(\sigma)\dag} a_{\bf k}^{(\sigma)} |\phi\rangle,
\el
where the first term is the vacuum part,
the second term is the photon part,
and the spectral energy density and pressure in de Sitter space is
\bl \label{rhopTRvac}
\rho^{TR}_k  & =   3 p^{TR}_k
 = \frac{k^3}{2\pi^2 a^4}
  \Big[ |f_k^{(1)'}(\tau)|^2 + k^2 |f_k^{(1)}(\tau)|^2  \Big]
  \nn \\
&  = \frac{k^4}{2 \pi^2 a^4}  ,
\el
where the transverse mode $f_k^{(1)}$ is given by \eqref{f12m0solu}.
If the photon  part during de Sitter inflation
is in thermal equilibrium approximately,
the photon number distribution will be described by
$\langle \phi|a_{\bf k}^{(\sigma)\dag} a_{\bf k}^{(\sigma)} |\phi\rangle
  \propto 1/(e^{k/T} -1)$,
and the integration over $k$ yields
the photon part of transverse  energy density
\bl \label{radiation}
\int\frac{d k}{k}   \rho^{TR}_k
   \sum_{\sigma=1,2}
  \langle \phi|a_{\bf k}^{(\sigma)\dag}
               a_{\bf k}^{(\sigma)} |\phi\rangle
  = \frac{\pi^2}{15} \Big( \frac{T}{a(\tau)} \Big)^4 ,
\el
which is convergent, and diluting as $a^{-4}$ with the cosmic expansion.
We are more interested in the  vacuum part.
The transverse vacuum spectral stress tensor \eqref{rhopTRvac}
has only one UV divergent $ k^4$ term,
which is similar to that
in the Minkowski spacetime  (see \eqref{transvrho} in Appendix B).
Since the solution \eqref{f12m0solu} of $B_i$
holds for a general RW spacetime,
so does  the transverse stress tensor  \eqref{rhopTRvac},
which
also respects  the conservation law in a general RW spacetime
\be
\rho_k^{TR} \, ' +3 \frac{a'}{a}   (\rho_k^{TR}+p_k^{TR}) =0.
\ee

The LT  stress tensor should be removed
since the longitudinal and temporal fields
 are not  radiative  dynamical degree of freedom.
This is conventionally  implemented by adopting
the GB physical state \cite{Gupta1977,Gupta1950,ItzyksonZuber}.
For the longitudinal and temporal fields,
the GB physical states $|\psi\rangle$ are defined as the following.
The positive frequency  part of
the temporal canonical momentum operator $\pi_A^0$ of \eqref{N31}
annihilates the state $|\psi\rangle$,
\bl
& \pi_A^{0  (+)}|\psi\rangle=0 ,
~~~ \rightarrow ~~~
( c_2a^{(0)}_{\bf k}+m_2a^{(3)}_{\bf k} )|\psi\rangle=0 .
 \label{N46}
\el
This GB condition on the physical state
is weaker than the Lorenz condition $(\nabla^\nu A_\nu=0)$
on the field operators.
By the choice \eqref{allowc12}, $c_2=-m_2$,
\eqref{N46}  can be written as
\bl
[ a^{(0)}_{\bf k} - a^{(3)}_{\bf k} ]|\psi\rangle=0 ,
\label{chN46}
\el
which also implies
\bl  \label{LTGBcondition}
\langle\psi|a^{(0)\dag}_{\bf k} a^{(0)}_{\bf k} |\psi\rangle
   =\langle\psi|a^{(3)\dag}_{\bf k} a^{(3)}_{\bf k} |\psi\rangle,
\el
ie,  the number of temporal
and longitude photons are equal in the GB physical state.
Together with the transverse state $|\phi \rangle$,
the complete state of the Maxwell field
can be denoted as a direct product
$|\phi,\psi\rangle=|\phi\rangle\otimes|\psi\rangle$.
It is known that the GB condition \eqref{N46}
may not hold for a general RW spacetime \cite{Higuchi1990},
where the positive frequency modes in the asymptotic in-region may
evolve into a combination of positive and negative frequency modes
in the asymptotic out-region.
This  generally happens when the cosmic expansion
consists of several stages of power-law expansion
 \cite{WangZhangChen2016,ZhangWangJCAP2018}.
However, during the de Sitter expansion \eqref{metric} \eqref{deSittera},
the positive frequency modes
\eqref{H12} $\sim$ \eqref{pi0des}
remain $\propto e^{-i k \tau }$ for the whole range of $\tau$,
so that the GB condition \eqref{N46} can be imposed consistently.

The expectation of the LT stress tensor in the  GB physical state is
\bl
\langle\psi|\rho^{LT}|\psi\rangle = 3  \langle\psi|p^{LT}|\psi\rangle
=\frac{1}{2} a^{-4}
         \langle\psi|\partial_i\pi_A\partial^i \pi_A |\psi\rangle .
  \label{pipi}
\el
Substituting the operator $\pi_A$ of \eqref{N33} into the above gives
\bl
\langle\psi|\rho^{LT}|\psi\rangle
& = 3 \langle\psi|p^{LT}|\psi\rangle
=      \int\rho^{LT}_{k} \frac{d k}{k } ,
   \label{rhoLT}
\el
where
\bl
\rho^{LT}_{k}
&=  \frac{k^5}{4 \pi^2  a^4}
 \bigg(2\langle\psi| a^{(0)\dag}_{\bf k} a^{(0)}_{\bf k}
 |\psi\rangle |\pi_{A1k}|^2
+2\langle\psi| a^{(3)\dag}_{\bf k} a^{(3)}_{\bf k}  |\psi\rangle |\pi_{A2k}|^2
  - |\pi_{A1k}|^2   + |\pi_{A2k}|^2 \bigg)
 \nn \\
& +  \frac{k^5}{4 \pi^2  a^4}
 \bigg(   2 \langle\psi| a^{(3)\dag}_{\bf k} a^{(0)}_{\bf k}
  |\psi \rangle \pi_{A2k}^*\pi_{A1k}
 + 2 \langle\psi| a^{(0)\dag}_{\bf k} a^{(3)}_{\bf k}
  |\psi\rangle  \pi_{A1k}^*\pi_{A2k}    \bigg)
\nn \\
& +   \frac{k^5}{4 \pi^2  a^4}
\bigg(\langle\psi|a^{(0)}_{\bf k} a^{(0)}_{-\bf k} |\psi\rangle\pi_{A1k}^2
+\langle\psi|a^{(3)}_{\bf k} a^{(0)}_{-\bf k} |\psi\rangle\pi_{A2k}\pi_{A1k}
\nn \\
& ~~~~~~~~~~
 +\langle\psi|a^{(0)}_{\bf k} a^{(3)}_{-\bf k} |\psi\rangle\pi_{A1k}\pi_{A2k}
+\langle\psi|a^{(3)}_{\bf k} a^{(3)}_{-\bf k}|\psi\rangle\pi_{A2k}^2
  +h.c.\bigg) .\label{LTrho}
\el
Applying  the GB condition \eqref{N46} \eqref{LTGBcondition}
and the mode relation  \eqref{relcona} with $c_2=-m_2$,
we find that the longitudinal and temporal contributions cancel each other,
and \eqref{LTrho}  becomes
\bl \label{masslessLT}
\rho_k^{LT} = 3p_k^{LT} =0 ,
\el
including the photon and vacuum parts.
Thus,
the LT stress tensor is vanishing in the GB state even before regularization.
This result is independent of $\zeta$.
The longitudinal-temporal cancelation occurs in the GB state
as long as the modes $\pi_{A 1k}$ and  $\pi_{A 2k}$
satisfy the relation  \eqref{relcona},
regardless the concrete functions  $\pi_{A 1k}$ and  $\pi_{A 2k}$.
We have also checked that
the LT stress tensor is zero also
for the radiation dominant stage ($a \propto \tau$).
So, it might be  expected that
the LT stress tensor will be zero for a general
 power-law expansion with $a \propto \tau^n$.
But this may not hold in a general RW spacetime
consisting of several stages of power-law expansion.

More interesting is
the GF  stress tensor which is less studied in literature.
The expressions \eqref{rhoGFml}  \eqref{GFpreml}
 in the GB physical state give
\bl
\langle \psi |\rho^{GF }|\psi\rangle
 & =  \frac{1}{a^4}\langle \psi |
  \Big(- \frac1{2} \zeta  (\pi_A^0)^2
  -   A_0 \big( \partial_0 \pi_A^{0} - D \pi_A^0 \big)
  -  A_{,j} \pi^0_{A \, , \,  j}  \Big)|\psi\rangle ,
   \label{gfPIPI}
  \\
\langle \psi |p^{GF }|\psi\rangle
& =  \frac{1}{a^4}\langle \psi |\Big( \frac1{2} \zeta  (\pi_A^0)^2
  -   A_0 \big( \partial_0 \pi_A^{0} - D \pi_A^0 \big)
  + \frac13   A_{,j} \pi^0_{A \, , \,  j}  \Big)|\psi\rangle .
  \label{PIPIPI}
\el
It can be shown that the expectation value
 $\langle \psi |(\pi_A^0)^2|\psi\rangle =0$ in the GB physical state,
so \eqref{gfPIPI} \eqref{PIPIPI} reduce to
\bl
\langle \psi |\rho^{GF }|\psi\rangle
& =  \frac{1}{a^4}\langle \psi |\Big(
  -   A_0 \big( \partial_0 \pi_A^{0} - D \pi_A^0 \big)
  -  A_{,j} \pi^0_{A \, , \,  j}  \Big)|\psi\rangle  ,
  \label{PIPIPI2}
\\
\langle \psi |p^{GF }|\psi\rangle
& =  \frac{1}{a^4}\langle \psi |\Big(
  -   A_0 \big( \partial_0 \pi_A^{0} - D \pi_A^0 \big)
  + \frac13   A_{,j} \pi^0_{A \, , \,  j}  \Big)|\psi\rangle .
   \label{preIPI}
\el
Substituting  the operators   \eqref{N26} \eqref{N25} \eqref{N31}
into \eqref{PIPIPI2},
using the commutators \eqref{commaa} \eqref{commaa00},
 the mode relation \eqref{relcon},
the coefficient constraint \eqref{N44},
and the GB condition \eqref{N46},
we obtain
\bl
\langle \psi | \rho^{GF }|\psi\rangle
&=  \int \rho^{GF }_k  \frac{d k}{k}  ,  \label{rhogfgf}
\\
\langle \psi | p^{GF }|\psi\rangle
&=  \int p^{GF }_k  \frac{d k}{k}  ,  \label{pressgfgf}
\el
where the GF spectral energy density and   pressure are
\bl
\rho^{GF}_{k}
=   &  \frac{k^3 }{ 2\pi^2  a^4}  \Bigg[
   \langle \psi |a^{(0)\dag}_{\bf k} a^{(0)}_{\bf k} |\psi\rangle
 \Big( ( \frac{c_2}{m_2}  A_{02k} -A_{01k} )
 ( \partial_0 - D )\pi^{0*}_{A1k}
    + k^2 ( \frac{c_2}{m_2}  A_{2k} - A_{1k})  \pi^{0*}_{A1k}
    \Big) \nn
\\
&  +\langle \psi |a^{(3)\dag}_{\bf k} a^{(3)}_{\bf k} |\psi\rangle
  \Big( ( \frac{m_2}{c_2}  A_{01k} -  A_{02k} ) ( \partial_0 - D )\pi^{0*}_{A2k}
     + k^2 ( \frac{m_2}{c_2}  A_{1k} -A_{2k} ) \pi^{0*}_{A2k}
    \Big)
     \nn
\\
& ~~ + \Big( A_{01k}( \partial_0 - D)\pi^{0*}_{A1k}
                   -  A_{02k}( \partial_0 - D )\pi^{0*}_{A2k}
    + k^2 A_{1k} \pi^{0*}_{A1k}
    - k^2 A_{2k} \pi^{0*}_{A2k} \Big)  \Bigg] ,
    \label{uclu}
\el
\bl
p^{GF}_{k}
 =  & \frac{k^3 }{ 2\pi^2  a^4}  \Bigg[
   \langle \psi |a^{(0)\dag}_{\bf k} a^{(0)}_{\bf k} |\psi\rangle
 \Big(   (  \frac{c_2}{m_2}  A_{02k} -A_{01k} ) ( \partial_0 - D )\pi^{0*}_{A1k}
    -\frac13 k^2 ( \frac{c_2}{m_2}  A_{2k} - A_{1k})  \pi^{0*}_{A1k}
    \Big) \nn
\\
&  +\langle \psi |a^{(3)\dag}_{\bf k} a^{(3)}_{\bf k} |\psi\rangle
  \Big( ( \frac{m_2}{c_2}  A_{01k} -  A_{02k} ) ( \partial_0 - D )\pi^{0*}_{A2k}
     -\frac13 k^2 ( \frac{m_2}{c_2}  A_{1k} -A_{2k} ) \pi^{0*}_{A2k}
    \Big)
     \nn
\\
& ~~ + \Big( A_{01k}( \partial_0 - D)\pi^{0*}_{A1k}
                   -  A_{02k}( \partial_0 - D )\pi^{0*}_{A2k}
    - \frac13 k^2A_{1k}\pi^{0*}_{A1k}
    + \frac13 k^2A_{2k}\pi^{0*}_{A2k} \Big)  \Bigg] ,
     \label{puclu}
\el
each   consisting   of three contributions:
  the temporal photons, the longitudinal photons,
and the vacuum.
Substituting the modes $A_{1k}$,  $ A_{2k}$, $A_{01k}$, $ A_{02k}$,
of  \eqref{N28} $\sim$ \eqref{N29}
and the modes $\pi_{A1}^{0}$, $\pi_{A2}^{0}$
of \eqref{piA1slo} \eqref{N32slo} into \eqref{uclu} \eqref{puclu},
we do  lengthy calculation.
As we notice,
the inhomogeneous parts of $A_{0}$  cancel
in each of the following combinations
$(  \frac{c_2}{m_2}  A_{02k} -A_{01k} )$,
$( \frac{m_2}{c_2}  A_{01k} -  A_{02k} )$,
$A_{01k}( \partial_0 - D)\pi^{0*}_{A1k}
    -  A_{02k}( \partial_0 - D )\pi^{0*}_{A2k}$,
and similarly,
the inhomogeneous parts of $A$ cancel in the following
$(\frac{c_2}{m_2}  A_{2k} -A_{1k} )$,
$(\frac{m_2}{c_2}  A_{1k} -  A_{2k} )$,
$(A_{1k} \pi^{0*}_{A1k}
    -  A_{2k} \pi^{0*}_{A2k})$.
So, only the homogeneous parts
contribute to \eqref{uclu}  \eqref{puclu}, yielding
\bl
\rho^{GF}_k   =  &  \frac{k^4  }{ (2\pi^2)  a^4}
\Big[\Big( \langle\psi|a^{(3)\dag}_{\bf k} a^{(3)}_{\bf k} |\psi\rangle
 - \langle\psi|a^{(0)\dag}_{\bf k} a^{(0)}_{\bf k} |\psi\rangle \Big)
    \Big( 1 + \frac{1}{ 2 k^2 \tau ^2} \Big)  \Big]
     \nn \\
&       +  \frac{k^4 }{ 2\pi^2  a^4}
        \Big(1 +  \frac{1}{2 k^2 \tau^2} \Big) ,
        \label{N197}
\\
p_k^{GF}  =  &  \frac{k^4 }{ (2\pi^2)  a^4}  \frac{1}{3}
\Big[
  \Big( \langle\psi|a^{(3)\dag}_{\bf k} a^{(3)}_{\bf k} |\psi\rangle
 - \langle\psi|a^{(0)\dag}_{\bf k} a^{(0)}_{\bf k} |\psi\rangle \Big)
    \Big( 1 - \frac{1}{ 2 k^2 \tau ^2} \Big)
                 \Big]
                 \nn \\
&  +   \frac{k^4 }{ (2\pi^2)  a^4}   \frac{1}{3}
  \Big( 1  - \frac{1}{ 2 k^2 \tau^2} \Big) .
      \label{opgfp}
\el
By
$\langle \psi |a^{(0)\dag}_{\bf k} a^{(0)}_{\bf k} |\psi\rangle
= \langle \psi |a^{(3)\dag}_{\bf k} a^{(3)}_{\bf k} |\psi\rangle$,
the longitudinal and temporal photons cancel each other,
only the vacuum part remains
\bl
\rho^{GF}_{k} & =  \frac{k^4 }{ 2\pi^2  a^4}
        \Big(1 +  \frac{1}{2 k^2 \tau^2} \Big)
        \label{rhogf97} ,
\\
p^{GF}_{k}  & =    \frac{k^4 }{ 2\pi^2  a^4}   \frac{1}{3}
  \Big(  1   -  \frac{1}{ 2 k^2 \tau^2}  \Big)  .
     \label{pressfg}
\el
This GF vacuum part  is  independent of $\zeta$ too,
because the $\zeta$-dependent, inhomogeneous parts
of the $k$-modes of $A$ and $A_0$ have canceled.
The GF vacuum stress tensor also respects the conservation law
\be
\rho^{GF}_k \, ' +3\frac{a'}{a} (\rho^{GF}_k +p^{GF}_k )=0 ,
\ee
and but contributes a nonzero trace
\bl\label{vevtrace}
- \rho^{GF}_k   + 3 p^{GF}_k =
 - \frac{k^4}{2\pi^2a^4}  \frac{1}{ ( k^2 \tau ^2)} \ne 0 .
\el
The form of  \eqref{rhogf97} \eqref{pressfg}
is the same as  twice the vacuum stress tensor
 of the minimally-coupling massless scalar field
\cite{ZhangYeWang2020,ZhangWangYe2020}.
It contains two UV divergent terms:
the $ k^4$ term is dominant
and corresponds to the UV divergence in the Minkowski spacetime
(see \eqref{GFenergymk} in Appendix B),
and the $k^2$ term   reflects  the effect of the cosmic expansion
and is absent in the Minkowski spacetime.

The transverse stress tensor and the LT stress tensor are invariant
under the residual gauge transformation
even at the classical level.
Now we examine the behavior
of the GF vacuum stress tensor \eqref{rhogf97} \eqref{pressfg}
under the quantum residual gauge transformation.
Firstly, according to \eqref{quantumtraf}, $c_1'\ne 0$ and $ m_1'\ne 0$,
the homogeneous part of $A$ and $A_0$
will not be transformed  to zero
under the quantum residual gauge transformation.
As a result,
the vacuum GF stress tensor will not be transformed zero
since it is contributed by the homogeneous part.
More than that,
 the GF stress tensor in the GB state is actually invariant
under the quantum residual gauge transformation.
This fact can be shown by a direct calculation
of the variation of the GF spectral stress tensor \eqref{uclu} \eqref{puclu}
\bl
\delta \rho ^{GF}_{k}
 = &  \frac{k^3}{2\pi^2a^4} \big(c_2^*C -  m_2^* M \big) i H
  \Big[ \Big( \langle\psi|a^{(3)\dag}_{\bf k} a^{(3)}_{\bf k} |\psi\rangle
 - \langle\psi|a^{(0)\dag}_{\bf k} a^{(0)}_{\bf k} |\psi\rangle \Big)
    \Big( 1 + \frac{1}{ 2 k^2 \tau ^2} \Big)
   \nn \\
&  +  \big(1+\frac{1}{2k^2\tau^2} \big) \Big] ,
\\
\delta p^{GF}_{k}
 =  & \frac{k^3}{2\pi^2a^4}  \frac13
  \big( c_2^* C-  m_2^* M \big)  i H
  \Big[ \Big( \langle\psi|a^{(3)\dag}_{\bf k} a^{(3)}_{\bf k} |\psi\rangle
 - \langle\psi|a^{(0)\dag}_{\bf k} a^{(0)}_{\bf k} |\psi\rangle \Big)
    \Big( 1 - \frac{1}{ 2 k^2 \tau ^2} \Big)
    \nn \\
&    +  \big(1 - \frac{1}{2k^2\tau^2}\big) \Big].
\el
According to the constraint
$ m_2 M^* -c_2 C^*  =0$ of \eqref{forgaugetrans},
the above is vanishing
\be \label{gfgtr}
\delta \rho^{GF}_{k}=0,
~~ ~~~ \delta p^{GF}_{k}=0 .
\ee

\section{\bf The regularization of stress tensor
of Maxwell field in de Sitter space}

So far three parts of  the vacuum stress tensor have been derived
in de Sitter space.
The LT stress tensor \eqref{masslessLT} is zero  in the GB state,
 no need for regularization.
The transverse vacuum  stress tensor
and the GF vacuum stress tensor both contain UV divergences,
which need to be regularized as the following.

The transverse vacuum stress tensor \eqref{rhopTRvac}
has only one quartic $k^4$ divergent term,
so the 0th-order adiabatic regularization is sufficient
to remove the UV divergence
\cite{WangZhangChen2016,ZhangWangJCAP2018,
ZhangYeWang2020,ZhangWangYe2020,YeZhangWang2022}.
The equation of two transverse modes
is eq.\eqref{Bieq1mk}
and the exact solution is $f_k^{(\sigma)}(\tau)$ in \eqref{f12m0solu}.
The adiabatic transverse modes are the same for two polarizations $(\sigma=1,2)$,
given by the WKB solution of \eqref{Bieq1mk} as the following
\cite{ParkerFulling1974,FullingParkerHu1974,
WangZhangChen2016,ZhangWangJCAP2018,
ZhangWangYe2020,ZhangYeWang2020}
\be\label{Yvntr}
f_k(\tau)
  = (2W(\tau))^{-1/2} \exp \Big[  -i \int^{\tau} W(\tau')d\tau' \Big],
\ee
where the effective frequency is
\be\label{YequaWktr}
  W(\tau)     = \Big[  \omega^2
-\frac12 \l( \frac{ W  '' }{ W}
- \frac32 \big( \frac{W  '}{W} \big)^2 \r)  \Big]^{1/2} ,
\ee
which will be solved iteratively.
The 0th-order frequency and mode are
\bl
W_{0th} & = \omega = k   ,
\\
f_{k\, 0th} (\tau) &  =\frac{1}{\sqrt{2 k}}e^{- ik \tau}
= f_k^{(\sigma)} . \label{ad0f}
\el
In fact,  all adiabatic orders for the transverse modes are  the same
\bl
W_{0th} & = W_{2nd}  =W_{4th}  = ... =k ,
\\
f_{k\, 0th} (\tau) & = f_{k\, 2nd} (\tau) = f_{k\, 4th} (\tau)
= ... =   f_k^{(\sigma)} ,
\label{admf}
\el
like a conformally-coupling massless scalar field
\cite{ZhangYeWang2020,ZhangWangYe2020}.
Substituting the 0th-order  mode $f_{k\, 0th}$ of \eqref{ad0f}
into \eqref{rhopTRvac} to replace $f_k^{(1)}$ yields
\bl \label{adrrhtr}
\rho^{TR }_{k \, \, 0th}  & = 3  p^{TR}_{k  \, \, 0th}  =
  \frac{k^3}{2\pi^2 a^4}
  \Big[ |f_{k\, 0th}'(\tau)|^2 + k^2 |f_{k\, 0th}(\tau)|^2  \Big]
   = \frac{k^4}{2 \pi^2 a^4}
\nn \\
&  = \rho^{TR}_{k}  = 3p^{TR}_{k} ,
\el
ie, the 0th-order adiabatic subtraction term
for the transverse spectral stress tensor
is just equal to the exact spectral stress tensor  \eqref{rhopTRvac}.
Hence, by subtraction,
the 0th-order regularized transverse vacuum
spectral stress tensor is vanishing
\bl \label{adrrhtrho}
\rho^{TR}_{k\, reg} \equiv \rho^{TR}_{k} -\rho^{TR }_{k \, \, 0th}   =0,
\\
p^{TR}_{k\, reg} \equiv p^{TR}_{k} - p^{TR }_{k \, \, 0th}  =0.
  \label{adrrhtp}
\el
The results \eqref{adrrhtr} $\sim$ \eqref{adrrhtp} hold
also for a general RW spacetime.
This is because  $B_i$ of   \eqref{f12m0solu}
and  its adiabatic modes \eqref{admf}
hold for a general RW spacetime \cite{ZhangWangYe2020}.

The GF  vacuum stress tensor  \eqref{rhogf97} \eqref{pressfg}
has  the $k^4$ and $k^2$ divergent terms,
so the 2nd-order adiabatic regularization is sufficient
to remove the UV divergences
\cite{WangZhangChen2016,ZhangWangJCAP2018,
ZhangYeWang2020,ZhangWangYe2020,YeZhangWang2022}.
To calculate the 2nd-order adiabatic subtraction terms
 of the stress tensor,
we need also  respectively the 2nd-order adiabatic modes of
$\pi_{Ak}^0$, $A_{k}$ and  $A_{0k}$.

The equation of rescaled $\bar \pi_A^{0} $ is given by \eqref{barpi0eq}
and the  solution  is given by \eqref{Y0sol}.
The WKB solution of \eqref{barpi0eq} is
\be\label{Yvntr}
\bar \pi_{A \,  nth}^{0}
  = (2W(\tau))^{-1/2} \exp \Big[  -i \int^{\tau} W(\tau')d\tau' \Big],
\ee
where the effective frequency is
\be\label{YequaWktr}
  W(\tau)     = \Big[  \omega^2 -\frac{2}{\tau^2}
-\frac12 \l( \frac{ W  '' }{ W}
- \frac32 \big( \frac{W  '}{W} \big)^2 \r)  \Big]^{1/2} ,
\ee
which will be solved iteratively.
The 0th-order is $W_{0th}  = \omega = k $,
and the 2nd-order and the higher orders  are found
\bl
W_{2nd}  =W_{4th}   = ...=  k -\frac{1}{k \tau^2}   ,
\el
so the 2nd-order and all higher order adiabatic modes  are
the same, and given by
\bl
\bar \pi_{A \,  2nd}^{0} & = \bar \pi_{A \,  4th}^{0}
 =  \bar \pi_{A \,  6th}^{0} =  ...
\nn  \\
& = \frac{1}{\sqrt{2 k} } \big(1 + \frac12 \frac{1}{(k\tau)^2 } \big)
\exp \Big[  -i k \int^{\tau} \big(1 - \frac{1}{(k\tau')^2}\big) d\tau' \Big]
\nn \\
& \simeq  \frac{1}{\sqrt{2 k} }
\big( 1 - \frac{i}{k\tau}   \big)   e^{-i k \tau} ,
\el
which is equal to the exact mode $ \bar \pi_{A}^{0}$ in \eqref{Y0sol}.
Multiplying by $a(\tau)$, one has  $\pi_{A\, 2nd}^{0}=\pi_{A}^{0}$,
ie,  the 2nd and higher order adiabatic modes are equal to
the exact modes \eqref{piA1slo}  \eqref{N32slo}.

The WKB approximation of $A$ and $A^0$
can be derived,  in principle,
  from their fourth order differential equations,
but the calculation will be more involved.
Actually we can directly  get their 2nd-order adiabatic modes
from high $k$ expansions of the exact modes
 \eqref{N28}  $\sim$  \eqref{N29}.
Moreover, as  mentioned earlier,
the inhomogeneous part of  $A$ and $A_0$ do not contribute to
the GF stress tensor,
so we need only  the homogeneous parts of
\eqref{N28} $\sim$ \eqref{N29} as  the following
\bl
A_{1k } &  =  c_1\frac{1}{a(\tau)}  \frac{i}{k}
 (1-\frac{i}{k\tau })\frac{1}{\sqrt{2 k}} e^{-i k \tau } ,
    \label{SSS7}
\\
A_{2k }  & =   m_1\frac{1}{a(\tau)}
     \frac{i}{k}  (1-\frac{i}{k\tau })\frac{1}{\sqrt{2 k}} e^{-i k \tau } ,
   \label{SSS9}
\el
which are of the 2nd adiabatic order already, and
\bl
A_{01k }
& = c_1 \frac{1}{a(\tau)} \frac{1}{\sqrt{2k}}  e^{-ik\tau} ,\label{SSS8}
\\
A_{02k }
& = m_1 \frac{1}{a(\tau)} \frac{1}{\sqrt{2k}}  e^{-ik\tau}  ,\label{SSS10}
\el
which are of the 0th adiabatic order,
and are also equal to all higher order homogeneous modes.
(Similarly,  for $\pi_A$,
the adiabatic modes of all orders are equal to the exact mode \eqref{pides}.
Here we shall not need these for regularization.)
Substituting these adiabatic modes
into  the expressions \eqref{uclu} \eqref{puclu}
to replace $\pi^{0}_{A1k}$, $\pi^{0}_{A2k}$,
$A_{1k }$,  $A_{2k }$,  $A_{01k }$,  $A_{02k }$,
we obtain
\bl
\rho^{GF}_{k\, 2nd}
  & = \frac{k^4}{2\pi^2  a^4} \Big(1 + \frac{1}{2 k^2 \tau^2} \Big)
    = \rho^{GF}_{k},
\\
p^{GF}_{k\, 2nd}
  & =  \frac{1}{3} \frac{k^4 }{ 2\pi^2   a^4}
  \Big( 1 - \frac{1}{2 k^2 \tau^2} \Big) = p^{GF}_{k},
\el
As expected,   the 2nd-order adiabatic subtraction term for
the GF spectral stress tensor
is equal to the exact GF spectral stress tensor.
By subtraction,
  the 2nd-order regularized GF vacuum  stress tensor is zero,
\bl
\rho^{GF}_{k\, reg} \equiv \rho^{GF}_{k} -\rho^{GF}_{k\, 2nd} =0 ,
   \label{GFrhoreg}
\\
p^{GF}_{k\, reg} \equiv p^{GF}_{k} - p^{GF}_{k\, 2nd} =0 ,
           \label{GFpreg}
\el
and  the regularized trace is also zero
\be
-\rho^{GF}_{k\, reg} +3 p^{GF}_{k\, reg} =0.
\ee
So, there is no need to introduce a ghost field
to cancel the  vanishing GF vacuum stress tensor
 \eqref{GFrhoreg} \eqref{GFpreg},
and this vanishing vacuum stress tensor
can not be a candidate for
the cosmological constant \cite{JimenezMaroto2009,JimenezMaroto2010}.
(Instead,  the regularized vacuum stress tensor
of a  massive scalar field,
either minimally-  or conformally-coupling,
does  give rise to the cosmological constant
\cite{ZhangYeWang2020,YeZhangWang2022}).
Putting the three parts  together,
the total regularized vacuum stress tensor of
Maxwell field with a general GF  term is zero,
\be
\rho_{reg} = p_{reg} =0,
\ee
and there is no trace anomaly.
This result is independent of $\zeta$,
and also invariant under the quantum residual gauge transformation.
Ref.\cite{AdlerLiebermanNg1977} adopted
the point-splitting regularization
\cite{DeWittBrehme1960,Christensen1976,YeZhangWang2022},
and also arrived at the zero vacuum stress tensor
of the Maxwell field in the Feynman gauge,
at the price of introducing
a ghost field to cancel the GF stress tensor.
The trace anomaly has been regarded as a consensus since 70's,
nevertheless our calculation shows no trace anomaly
for the Maxwell field.
Ref.\cite{WaldPRD1978,AdlerLieberman1978} claimed
the trace anomaly under the assumption that the Green's function
contains a boundary term $w(x,x')$
which is unsymmetric in  $(x,x')$.
But, as weshow,
the exact Green's function \eqref{sl}  \eqref{slc}  in de Sitter space
do not contain such an unsymmetric boundary term
\cite{YeZhangWang2022,ZhangYeWang2020,ZhangWangYe2020}.

\section{Conclusion and Discussions}

We have studied the Maxwell field with a general gauge fixing term
in de Sitter space.
All the four components $A_\mu$ are formally treated
as independent variables,
and no Lorenz condition is imposed.
The introduction of the GF  term
restricts the gauge invariance of the Maxwell field
down to a residual gauge invariance given by \eqref{soltheta}.
Furthermore, the covariant canonical quantization restricts further
the residual gauge invariance
down to the quantum residual gauge invariance
 specified by eq.\eqref{reggtf}.

The transverse components $B_i$  are separated from other components,
independent of the gauge fixing constant $\zeta$,
and represent real dynamical degrees of freedom,
and their equation \eqref{Bieq1mk} and solution \eqref{f12m0solu}
hold for a general RW spacetime including de Sitter space.
The transverse stress tensor
\eqref{rhomasslessdeSitter}
consists of the particle  parts \eqref{radiation}
and the vacuum part \eqref{rhopTRvac}
with a UV divergent term $\propto k^4$.

The longitudinal and temporal components
$A$ and $A_0$ are mixed up
in the $\zeta$-dependent equations (\ref{0aA}) (\ref{Aa0}).
We have obtained their solutions \eqref{H12}  \eqref{A0bar}
in two different ways.
In particular, in the second way,
via the  inhomogeneous equations \eqref{AEu2} \eqref{A0epi},
the nontrivial structure of the solutions $A$ and $A_0$
is revealed,
each being a sum of the homogeneous and inhomogeneous solutions.
The canonical momenta are contributed
only by the inhomogeneous solutions of  $A$ and $A_0$,
and only the homogeneous parts will vary
under the residual gauge transformation \eqref{Atrsf} \eqref{A0trsf}.
For a consistent covariant canonical quantization,
both the homogeneous and inhomogeneous $k$-modes of $A$ and $A_0$
need to be present in the operator expansions.
Moreover,
the homogeneous $k$-modes of $A$ and $A_0$ will not go vanishing
under the quantum residual gauge transformation.
The LT stress tensor \eqref{pipi}
is independent of $\zeta$,
and invariant under the quantum residual gauge transformation.
And its expectation \eqref{masslessLT} is zero in the GB  physical state
due to the longitudinal and temporal cancelation.

More interesting
is the GF  stress tensor,
which is less studied in literature.
At the classical level
the GF stress tensor \eqref{rhoGFml} \eqref{GFpreml}
depends upon $\zeta$,
nevertheless,
its expectation value \eqref{N197} \eqref{opgfp}
in the GB physical state is independent of $\zeta$,
and also is invariant under
the quantum residual gauge transformation.
Moreover,  its  particle part
is zero due to the longitudinal and temporal cancelation,
only the vacuum part \eqref{rhogf97} \eqref{pressfg} remains,
which contains two UV divergent terms,  $\propto k^4, k^2$,
and is equal to  twice the vacuum stress tensor
of the minimally-coupling massless scalar field.

To remove the UV divergences
of the vacuum stress tensor,
we have carried out the adiabatic regularization.
The transverse vacuum stress tensor becomes zero
under the 0th-order adiabatic regularization,
and, respectively, the GF vacuum stress tensor becomes zero
under the 2nd-order adiabatic regularization.
Thus,
there is no need to introduce a ghost field to cancel
the GF stress tensor,
and the vanishing vacuum GF stress tensor of Maxwell field
can not be a possible candidate for  the cosmological constant.
Instead, the regularized vacuum stress tensor
of  a (minimally- or conformally-coupling) massive scalar field
corresponds to the cosmological constant
that drives the de Sitter inflation
\cite{ZhangYeWang2020,YeZhangWang2022}.

In summary,  for the Maxwell field with a general GF term
in de Sitter space described by \eqref{metric} \eqref{deSittera},
the total regularized  vacuum stress tensor
in the GB state  is zero,
and only the photon part of the transverse  stress tensor \eqref{radiation}
remains,
and all the predicted physics
will be the same as that the Maxwell field
 without the GF  term.

We have also carried out
analogous calculations in the Minkowski spacetime,
attached in the Appendix B.
The outcome is similar to de Sitter space,
except that the GF vacuum stress tensors has
only one  $ k^4$ term,
which can be made zero by the normal ordering.

\

{\textbf{Acknowledgements}}

Y. Zhang is supported by
NSFC Grant No. 11675165,  11633001,  11961131007,
 and in part by National Key RD Program of China (2021YFC2203100).

\appendix
\numberwithin{equation}{section}

\section{ Green's functions for Maxwell field
in the Feynman gauge }

 Ref.\cite{DeWittBrehme1960} proposed the following  relations
(see also Ref.\cite{AdlerLiebermanNg1977} for the application)
\bl
G_{\nu\sigma^\prime }^{(1);\nu} & = -G_{S \, ,\sigma^\prime} ,
\label{00}
\\
G_{\nu \sigma^\prime}^{(1)\, ;\sigma^\prime} & = -G_{S \, ,\nu }  ,
\label{00'}
\el
where
\bl
G_{\nu\sigma^\prime}^{(1)}(x,x')&=\langle0|(A_{\nu}(x)A_{\sigma'}(x')
+A_{\sigma'}(x')A_{\nu}(x))|0\rangle,
\label{R2222}
\el
is the Hadamard type Green's function
for the  Maxwell field in the Feynman gauge ($\zeta=1$),
and
\bl
G_{S}(x,x') =\langle0|\phi(x)\phi(x')+\phi(x')\phi(x)|0\rangle ,
\el
is the Green's function for a minimally-coupling massless scalar field
where $\phi(x)$ is the scalar field operator.
Note that
$G_{\nu\sigma^\prime}^{(1)}(x,x')$ is not an ordinary tensor,
but a bi-vector at $x$ and at $x'$ respectively.
Similarly, $G_{S}(x,x')$ is a bi-scalar  at $x$ and at $x'$ respectively.
In the following
we  check the relation \eqref{00} in de Sitter space.

Write  the operator  $\phi$ as
\bl
\phi(x)=\int\frac{d^3 k}{(2\pi)^{\frac32}}\Big(a_{\vec{k}}
\phi_k(\tau) e^{i\bf{k}\cdot \bf{x}}   +
a_{\vec{k}}^\dag\phi_k(\tau)^* e^{-i\bf{k}\cdot \bf{x}}  \Big) ,
  \label{R21}
\el
where the $k$-mode of $\phi$
in de Sitter space is \cite{ZhangYeWang2020,ZhangWangYe2020}
\bl
\phi_{k}(\tau)=\frac{1}{a(\tau)}\frac{1}{\sqrt{2k}}
 (1-\frac{i }{k \tau  } )e^{-i k \tau  } .
      \label{R23}
\el
Simple calculation yields the Green function of the scalar field
\bl
G_{S}
&=\int\frac{d^3 k}{(2\pi)^{3}}\frac{H^2}{2k^3}
\Big( (-i+k \tau ) (i+k \tau')e^{-i k (\tau -\tau')}
      + c.c. \Big)
      e^{i\bf{k}\cdot (\bf{x}-\bf{x}')}   , \label{R299}
\el
After the $k$-integration, \eqref{R299} becomes \cite{YeZhangWang2022}
\bl \label{sl}
G_S (x,x')
 = -\frac{H^{2}}{8 \pi^{2}} \Big[ \frac{1}{\sigma}
+\ln (-\frac{2\tau\tau'}{\tau_0^2} \sigma) \Big]
\el
with $\sigma \equiv  \frac{1}{(2 \tau \tau')}
  [(\tau-\tau')^2-|{\bf x-x'|}^2 ]$
and  $\tau_0$ being a constant.
For the conformally-coupling massless scalar field
the Green function is
\bl \label{slc}
G  (x,x')
 = -\frac{H^{2}}{8 \pi^{2}}  \frac{1}{\sigma}  \,  ,
\el
which is relevant to the case
in Refs.\cite{AdlerLiebermanNg1977,AdlerLieberman1978,WaldPRD1978}.
Both \eqref{sl} and  \eqref{slc} are symmetric in $(x,x')$.
For an extension of \eqref{sl}
to vacuum states other than the Bunch-Davies vacuum state,
see Ref.\cite{Allen1985}.

The time and spatial derivatives of \eqref{R299} are
\bl
G_{S,0'}
&=\int\frac{d^3 k}{(2\pi)^{3}}\frac{(H\tau)(H\tau')}{2}
\Big(  (i+\frac{1}{k\tau} )  e^{-i k (\tau -\tau')}  + c.c. \Big)
e^{i\bf{k}\cdot (\bf{x}-\bf{x}')}  ,
\label{scgr}
\\
G_{S,i'}
&=-\int\frac{d^3 k}{(2\pi)^{3}}(ik_{i'})
\frac{H^2}{2k^3} \Big( (-i+k \tau ) (i+k \tau')e^{-i k (\tau -\tau')}
        + c.c.  \Big)
e^{i\bf{k}\cdot (\bf{x}-\bf{x}')}  .
\label{R51}
\el
From the solutions \eqref{f12m0solu} \eqref{H12}  \eqref{A0bar}
of $A_\mu(x)$  in de Sitter space,
we obtain each components of the Green's functions of Maxwell field
as the following
\bl
&G_{00'}^{(1)}=\int\frac{d^3k}{(2\pi)^{3}}
\frac{e^{-i k \left(\tau +\tau '\right)}}{6 k^3 \tau \tau'}
     \bigg[-i \zeta \big( \tau ^2 e^{2 i k \tau }+ \tau'\, ^2 e^{2 i k \tau '}\big)
    + k \zeta \tau \tau ' \big(\tau e^{2 i k \tau } + \tau' e^{2 i k \tau '}\big)
     \nn \\
&  - \zeta \tau^2 (i+k \tau') e^{2 i k \tau '}
   - \zeta \tau'\, ^2  (i+k \tau ) e^{2 i k \tau }
\nn
\\
& -i k^2 (-3+\zeta ) \tau ^2 \tau '^2 \Big(Ei(2 i k \tau)
+Ei(2 i k \tau ')+e^{2 i k (\tau +\tau ')}
  \big(Ei(-2 i k \tau )+Ei(-2 i k \tau ')\big)\Big) \bigg]
e^{-i\bf{k}\cdot (\bf{x}-\bf{x}')} ,\label{R10}
\el
\bl
G_{0i'}^{(1)}&=\int\frac{d^3k}{(2\pi)^{3}}\frac{ik_{i'}}{6 k^3 \tau}
\bigg[\bigg( e^{-i k \left(\tau +\tau '\right)} (1+ik \tau' )
\left(k^2\tau^2 (-3+\zeta ) Ei\left(2 i k \tau\right)
+e^{2 i k \tau} \zeta  \left(1-i k \tau \right)\right)\nn
\\
&~~~~~~~ + e^{i k \left(\tau +\tau '\right)}
e^{-2 i k \tau }  k^2 \tau ^2\left(3-2  \zeta +e^{-2 i k \tau' }
(-3+\zeta ) (1+ik \tau' ) Ei(2 i k \tau' )\right) \bigg)\nn
\\
&~~~ -\bigg( e^{i k \left(\tau +\tau '\right)}  (-1+i k \tau' )
\left(k^2\tau^2 (-3+\zeta ) Ei\left(-2 i k \tau\right)
+e^{-2 i k \tau} \zeta  \left(1+i k \tau\right)\right)\nn
\\
&~~~ + e^{-i k \left(\tau +\tau '\right)} e^{2 i k \tau}
k^2 \tau ^2 \left(-3 +2  \zeta +e^{2 i k \tau' } (-3+\zeta )
 (-1+ik \tau' ) Ei(-2 i k \tau' )\right)\bigg) \bigg]
e^{-i\bf{k}\cdot (\bf{x}-\bf{x}')} , \label{R35}
\el
\bl
G_{li'}^{(1)}
 & = \int\frac{d^3\vec{k}}{2k(2\pi)^3}(\delta_{li'}-\frac{k_lk_{i'}}{k^2})
  (e^{-i k (\tau-\tau')}+e^{-i k(\tau'-\tau)})
  e^{i\bf{k}\cdot (\bf{x}-\bf{x}')}
 \nn \\
&  +\int\frac{d^3k}{(2\pi)^{3}}  \frac{k_lk_{i'}}{6 k^5}e^{-i k (\tau +\tau')}
 \Big[ ik^2 (-i+k \tau ) (e^{2 i k \tau'} (-3+2 \zeta )-i (-3+\zeta )
  (-i+k \tau') Ei(2 i k \tau'))\nn
\\
&~~~~~~~~ - i k^2 e^{2 i k \tau'} (i+k\tau')
\left(-3+2 \zeta +i e^{2 i k \tau } (-3+\zeta ) (i+k \tau )Ei(-2 i k \tau)\right)
\nn
\\
&~~~~~~~~ + ik^2 (-i+k \tau' ) (e^{2 i k \tau} (-3+2 \zeta )
-i (-3+\zeta ) (-i+k \tau) Ei(2 i k \tau))\nn
\\
&~~~~~~~~ - i k^2 e^{2 i k \tau} (i+k\tau)
  \left(-3+2 \zeta +i e^{2 i k \tau' } (-3+\zeta ) (i+k \tau' )
  Ei(-2 i k \tau')\right)\Big]
  e^{i\bf{k}\cdot (\bf{x}-\bf{x}')} .
    \label{Gli}
\el
Each contains the exponential-integration function $Ei$.
In a homogeneous and isotropic RW spacetime, there is a symmetry
\[
G_{\mu {\nu'}}^{(1)} (x,x') = G_{ {\nu'} \mu}^{(1)} (x',x) ,
\]
so that
\[
 G_{i'0}^{(1)}(x',x)|_{x \, \leftrightarrow \, x'}
 = G_{0i'}^{(1)}(x,x') |_{x\, \leftrightarrow \, x'} .
\]
Since   $G^{(1)}_{\nu\sigma'}(x,x')$ is a  vector at the point $x$,
 the $0'$-component of the four divergence is calculated
\bl
G_{\nu0^\prime}^{(1) \, ; \,  \nu}
 & = g^{\mu\nu}G_{\nu0^\prime;\mu}^{(1)}
   = g^{\mu\nu}\Big(G_{\nu0^\prime,\mu}^{(1)}
      -\Gamma_{\nu\mu}^{\alpha}G_{\alpha0'}^{(1)}  \Big)
      \nn \\
& = a^{-2}\Big(-G_{00^\prime,0}^{(1)} +G_{i0^\prime,i}^{(1)}
           -2\frac{a'}a G_{00'}^{(1)}\Big)
\label{Gnu0nu}
\el
with $\Gamma_{00}^{0}=\frac{a'}{a}$,
$ \Gamma_{ij}^{0}=\delta_{ij} \frac{a'}{a}$,
$ \Gamma^{i}_{0j}=\frac{a'}{a}\delta_{ij}$.
Substituting \eqref{R10} \eqref{R35} \eqref{Gli}
with $\zeta=1$ into the above yields,
\bl
G_{\nu0^\prime}^{(1);\nu}=- G_{S,0'}  , \label{R31}
\el
where  $G_{S,0'}$ is given by \eqref{scgr},
and $Ei$ function has been canceled.
Similarly,   the $i'$-component of the four divergence is
\bl
G_{\nu i^\prime}^{(1);\nu}
& = \frac{1}{a(\tau)^2} \Big(-G^{(1)}_{0i',0}
-2\frac{a(\tau)'}{a(\tau)}G^{(1)}_{0 i'}+G^{(1)}_{l i',l} \Big).
\el
Calculation shows that
\bl
G_{\nu i^\prime}^{(1)\, ;\nu} =-G_{S,i'} \, ,
  \label{R52}
\el
where $G_{S,i'}$ is given by  \eqref{R51}.
So,  the relation \eqref{00} in the Feynman gauge  is verified.
Similarly,  \eqref{00'} can be also checked.
Note that  \eqref{00}  \eqref{00'} are not valid for a general $\zeta$.

\section{ Maxwell field  with a gauge fixing term
          in the Minkowski spacetime}

Although the Maxwell field in the Minkowski spacetime is well known,
the Maxwell field with a general GF term is nontrivial,
and has not been adequately reported in literature \cite{ItzyksonZuber}.
The procedure of calculation is analogous to that in de Sitter space.
In the following we shall report briefly the results.
Setting $D=0$ in \eqref{equaAm0zeta} gives the field equation
\bl \label{flatequaAm0zeta}
   \eta^{\sigma \rho} \partial_\sigma  \partial_\rho A_\nu &
      +\Big(  \frac{1}{\zeta} -1 \Big)
      \partial_\nu (\eta^{\rho\sigma} \partial_\sigma  A_\rho)
           =0 .
\el
Setting $D=0$ in  \eqref{0aA} \eqref{Aa0} gives
the following basic 2nd-order equations
\bl
-   \partial_0^2   A   -\frac{1}{\zeta} k^2  A
 + \big(1- \frac{1}{\zeta} \big)\partial_0  A_0  &  =0 ,
  \label{0aAmink}
\\
 - \frac{1}{\zeta}\partial_0^2  A_0  -k^2 A_0
     +  k^2  (1- \frac{1}{\zeta} )\partial_0  A & =0  ,
     \label{Aa0mink}
\el
where  $A_0$ and $A$ are mixed up,
and $B_i$ has the same equation and solution
as \eqref{Bieq1mk} \eqref{f12m0solu} in de Sitter space.
The decomposition is similar to \eqref{Aidecp} $\sim$ \eqref{pi0def},
but the temporal canonical momentum is
\bl   \label{pi0defmink}
\pi_A^0 & = - \frac{1}{\zeta} \big( \partial_0  A_0 + k^2  A \big) .
\el
Eqs.\eqref{piAeq1} \eqref{pi0Aeq} reduce to
$(\partial_0^2    +k^2 ) \pi_A =0$,  and
$( \partial_0^2   +k^2 ) \pi^0_A    =0$
in the Minkowski spacetime,
the positive frequency solutions are
\bl
\pi_A=d_1\frac{1}{\sqrt{2k}}e^{-i k t},
 ~~~~~~~\pi_{A}^0=d_2 (ik) \frac{1}{\sqrt{2k}}e^{-i k t}  ,\label{M449}
\el
where $d_1$ and $d_2$ are arbitrary coefficients.
By differentiation and combination
 of eqs.\eqref{0aAmink} \eqref{Aa0mink},
we obtain the 4th-order differential equations
\bl
(\partial_0^2 +k^2)^2 A    & =0 , \label{minkwA}    \\
 (\partial_0^2 +k^2 )^2 A_0 &  =0 , \label{4thAeq}
\el
which are separate,  and independent of $\zeta$,
unlike \eqref{A4ordereqSimpl0} \eqref{A04ordereqSimpl0}
in the de Sitter space.
The  positive  frequency   solutions of  \eqref{minkwA} \eqref{4thAeq} are
\bl
 A (\tau) & =  b_{} \frac{i}{k} \frac{1}{\sqrt{2k}} e^{ -  ik\tau }
   + c_{}   \frac{i}{k}  \frac{( 1+ 2 i k \tau  )}{\sqrt{2k}} e^{-i k \tau} ,
    \label{Aposmink}
  \\
 A_0 (\tau) & =  b_{0}  \frac{1}{\sqrt{2k}} e^{ -  ik\tau }
   + c_{0}    \frac{( 1+ 2 i k \tau  )}{\sqrt{2k}} e^{-i k \tau} ,
   \label{A0posmink}
\el
where $b,b_0,c,c_0$ are arbitrary constants.
Substituting  \eqref{Aposmink} \eqref{A0posmink}
into the basic equations \eqref{0aAmink} \eqref{Aa0mink}
to constrain  $(c,c_0,b,b_0)$,
we obtain, for  $\zeta \ne -1$,
\bl
A (\tau) & =  b  \frac{i}{k}  \frac{1}{\sqrt{2k}} e^{ -ik\tau }
   + \frac12  (b-b_0) \frac{\zeta-1}{\zeta+1}
     \frac{i}{k}  \frac{  (1+ 2 i k \tau  ) }{\sqrt{2k}} e^{ -ik\tau } ,
          \label{Aposminknf}
          \\
A_0 (\tau)
 & =  b \frac{1}{\sqrt{2k}} e^{- ik\tau}
 + (b-b_0)\Big(-1 + \frac12 \frac{\zeta-1}{\zeta+1} ( 1+2 i k \tau) \Big)
     \frac{1}{\sqrt{2k}} e^{ - ik\tau}  ,         \label{A0posminknf}
\\
\pi_A  & =   \frac{ 2 (b- b_0)}{\zeta +1} \frac{ 1}{ \sqrt{2 k}}e^{-i k \tau },
        \label{canmoma}
        \\
\pi^0_A  & =  \frac{2 (b- b_0) }{\zeta +1} (-ik)
 \frac{1}{ \sqrt{2 k}} e^{-i k \tau } ,
 \label{canmoma0}
\el
where the canonical momenta are contributed only by
the $(b-b_0)$-part of \eqref{Aposminknf} \eqref{A0posminknf}.
Similarly, for $\zeta\ne 1$,  we obtain
\bl
A (\tau) & =  b  \frac{i}{k}   \frac{1}{\sqrt{2k}} e^{ -ik\tau }
   + c   (1+ 2 i k \tau)  \frac{i}{k} \frac{1 }{\sqrt{2k}} e^{ -ik\tau },
            \label{Aposminknf2}
          \\
A_0 (\tau)
& =   b \frac{1}{\sqrt{2k}} e^{- ik\tau}
    +  c \Big( ( 1+2 i k \tau)- 2  \frac{\zeta+1}{\zeta-1} \Big)
           \frac{1}{\sqrt{2k}} e^{- ik\tau} ,
      \label{A0posminknf2}
\\
\pi_A  &  = c  \frac{ 4 }{\zeta - 1}  \frac{ 1}{ \sqrt{2 k}}e^{-i k \tau },
  \\
\pi^0_A  &  =  c \frac{ 4 }{\zeta - 1} (- ik)
 \frac{ 1}{ \sqrt{2 k}} e^{-i k \tau }      . \label{canmoma2}
\el
In the  Feynman gauge
eqs.\eqref{0aAmink} \eqref{Aa0mink} with $\zeta=1$  reduce  to
\bl
 (\partial_0^2 +k^2) A     & =0 , \label{feynmanminkwA}    \\
 (\partial_0^2 +k^2 )  A_0 &  =0 , \label{feynman4thAeq}
\el
which are  already  separated for  $A$ and $A_0$,
and the solutions \eqref{Aposminknf} -- \eqref{canmoma0} reduce to
\bl
A (\tau) & =  b  \frac{i}{k}  \frac{1}{\sqrt{2k}} e^{ -ik\tau }  ,
  ~~~~~
A_0 (\tau)  =  b_{0}  \frac{1}{\sqrt{2k}} e^{- ik\tau} ,
          \label{FeynA}
     \\
\pi_A  & =     (b- b_0) \frac{ 1}{ \sqrt{2 k}}e^{-i k \tau },
 ~~~~~
\pi^0_A     =   (b- b_0)  (-ik) \frac{1}{ \sqrt{2 k}} e^{-i k \tau } .
\el
The  Feynman  gauge is commonly used in the text books,
whereas a general  gauge is less addressed.

The solutions of $A$ and $A_0$ can be rederived by another way.
Setting $D=0$ in   \eqref{AEu2}  \eqref{A0epi}  leads to
the following inhomogeneous equations
\bl
\partial_0^2 A     +k^2 A
   & = \partial_0   \pi_A  - \zeta \pi_A^0  ,
    \label{M452}
\\
\partial_0^2  A_0   + k^2   A_0
     &  = -( k^2  \pi_A + \zeta \partial_0   \pi_A^0) .
  \label{M451}
\el
Since  $\pi_A$ and $\pi_A^0$ are known in \eqref{M449},
we get the solutions of \eqref{M452} \eqref{M451},
\bl
A & = b \frac{i}{k} \frac{1}{\sqrt{2k}}e^{-i k \tau }
    - \frac14 ( d_1 + d_2 \zeta ) \frac{i}{ k} \frac{(1+2 ik \tau )}{\sqrt{2k}}
         e^{-i k \tau },\label{M455}
         \\
A_0 & = b_0  \frac{1}{\sqrt{2k}}e^{-i k \tau }
  - \frac14 ( d_1 +  d_2    \zeta ) \frac{(1+2 i k \tau )}{\sqrt{2k} }
     e^{-i k \tau } .
  \label{M454}
\el
Substituting \eqref{M455} \eqref{M454} into \eqref{0aAmink} \eqref{Aa0mink}
leads to the constraints on the coefficients
\bl
(d_1 +  \zeta d_2) & = -  2 (b- b_0) \frac{(\zeta -1)}{(\zeta +1)}   ,
  ~~~ (\zeta\ne -1) ,
 \label{bb02}
   \\
 (b- b_0)  & = -\frac{(\zeta +1)}{2 (\zeta -1)} (d_1 +  \zeta d_2),
 ~~~ (\zeta\ne 1) .
\el
This  gives \eqref{Aposminknf} \eqref{A0posminknf} for $\zeta\ne -1$
and  \eqref{Aposminknf2} \eqref{A0posminknf2} for $\zeta\ne 1$,
respectively.

Given  these solutions,
we perform the canonical quantization for a general $\zeta$.
The quantization of the transverse fields $B_i$ is the same as
\eqref{expBi}--\eqref{BiWico} in de Sitter space.
The longitudinal and temporal operators $A$, $A_0$,
$\pi_A$ and $\pi_A^0$ for a general $\zeta$
are the same as   \eqref{N26} \eqref{N25} \eqref{N33} \eqref{N31},
but with the $k$-modes  (for $\zeta\ne 1$)
\bl
A_{1k} (\tau) & =  b_1  \frac{i}{k}   \frac{1}{\sqrt{2k}} e^{ -ik\tau }
      + c_1  \frac{i}{k} (1+ 2 i k \tau) \frac{1}{\sqrt{2k}} e^{ -ik\tau },
        \label{A1qt}
          \\
A_{2k} (\tau) & =  b_2  \frac{i}{k}   \frac{1}{\sqrt{2k}} e^{ -ik\tau }
        + c_2 \frac{i}{k} (1+ 2 i k \tau) \frac{1 }{\sqrt{2k}} e^{ -ik\tau },
      \label{A2qt}
          \\
A_{01 k} (\tau)  & =   b_1 \frac{1}{\sqrt{2k}} e^{- ik\tau}
                  +  c_1 \Big( ( 1+2 i k \tau)- 2  \frac{\zeta+1}{\zeta-1} \Big)
           \frac{1}{\sqrt{2k}} e^{- ik\tau} ,
         \label{A01qt}
      \\
A_{02 k} (\tau)  & =   b_2 \frac{1}{\sqrt{2k}} e^{- ik\tau}
                  +  c_2 \Big( ( 1+2 i k \tau)- 2  \frac{\zeta+1}{\zeta-1} \Big)
           \frac{1}{\sqrt{2k}} e^{- ik\tau} ,
         \label{A02qt}
\\
\pi_{A1k} &   =  c_1 \frac{4}{\zeta - 1} \frac{1}{ \sqrt{2 k}} e^{-i k \tau } ,
   \label{qtpiaa}
 \\
\pi_{A2k} &  = c_2  \frac{4}{\zeta - 1} \frac{1}{ \sqrt{2 k}} e^{-i k \tau } ,
            \label{qtpiaa2}
\\
 \label{qtpiaa0}
\pi^0_{A1k} & =  c_1 \frac{ 4 }{\zeta - 1} (- ik)
\frac{ 1}{ \sqrt{2 k}} e^{-i k \tau },
         \\
\pi^0_{A2k} & = c_2 \frac{ 4 }{\zeta - 1} (- ik)
\frac{ 1}{ \sqrt{2 k}} e^{-i k \tau } ,
            \label{qtpi0aa0}
\el
where $(b_1, c_1), (b_2, c_2)$ are two sets of  coefficients.
(In the Feynman gauge,
the conventional one-operator expansion of $A$ and $A_0$ will be used
since their equations are separated as
\eqref{feynmanminkwA} \eqref{feynman4thAeq}.)
We impose the covariant canonical  commutation relations
\bl
 [A^\mu(\tau,  {\bf x}),\pi_A^{\nu}(\tau, {\bf y})]
           =i\eta^{\mu \nu}\delta(\bf x- y) .
\label{commt}
\el
Substituting the operators
\eqref{N25}  \eqref{N26} \eqref{N31} \eqref{N33}
 into each $(\mu\nu)$ component of \eqref{commt},
using the commutation relations \eqref{commaamunu},
we obtain the following constraints
upon the coefficients (for $\zeta\ne 1$)
\bl
& ( c_{2*}b_2 -c_{1*}b_1 ) =  \frac{1}{4}(\zeta -1 ) ,
         \label{1212bbcc}
\\
&  |c_2 | ^2- |c_1 | ^2   = 0 ,  \\
& |b_2 |^2  - |b_1|^2  = 1. \label{bbcc}
\el
There are infinite many choices to satisfy the above constraints.
For instance, a simple choice is
$c_1  = - c_2= 1$,
$b_1  =  \frac{2}{\zeta-1}  -  \frac{\zeta-1}{8}$,
$b_2 =   - \frac{2}{\zeta-1} -    \frac{\zeta-1}{8}$.

The stress tensor is not actually used in the Minkowski spacetime
 since gravity is not considered.
Here, in analog to that in de Sitter space,
we calculate the stress tensor in Minkowski spacetime.
The transverse,  LT, and  GF  stress tensors are defined
similar to the expressions
\eqref{trstr} \eqref{longittempml}  \eqref{rhoGFml} \eqref{GFpreml}
with $a =1$ and $D=0$.
We list the main results.
The transverse stress tensor is
\bl \label{rhoMink}
\langle \phi|  \rho^{TR} |\phi \rangle
&  =  3 \langle \phi|p^{TR} |\phi \rangle
= \int^{\infty}_0   \rho^{TR}_k   \frac{d k}{k}
  + \int\frac{d k}{k}   \rho^{TR}_k
   \sum_{\sigma=1,2}
   \langle \phi | a_{\bf k}^{\dag(\sigma)} a_{\bf k}^{(\sigma)} |\phi \rangle ,
\el
where the transverse spectral stress tensor is
\bl
\rho^{TR}_k  =  \frac{k^3}{2\pi^2 }
  \Big[ |f_k^{(1)'}(\tau)|^2 + k^2 |f_k^{(1)}(\tau)|^2  \Big]
    = \frac{k^4}{2 \pi^2}
     =  3 p^{TR}_k .
    \label{transvrho}
\el
The first term of \eqref{rhoMink} is
the UV divergent  vacuum energy density  in Minkowski spacetime,
which is routinely removed by normal ordering of
the creation and annihilation operators.
The LT stress tensor in the GB state $|\psi\rangle$ is
\bl
\langle\psi|\rho^{LT}|\psi\rangle
&
 =3 \langle\psi| p^{LT}  |\psi\rangle
 = \frac{1}{2}
    \langle\psi|    \partial_i\pi_A\partial^i \pi_A  |\psi\rangle
    =      \int\rho^{LT}_{k} \frac{d k}{k } .
\el
where
\be
\rho^{LT}_k =3 p^{LT}_k  =0  .
\label{rhopkgf}
\ee
The  GF  stress tensor  in the GB state is
\be
\langle \psi | \rho^{GF }|\psi\rangle
=  \int \rho^{GF }_k  \frac{d k}{k}   ,
~~~~ \langle \psi | p^{GF }|\psi\rangle
=  \int p^{GF }_k  \frac{d k}{k}   ,
\label{rhogfgf}
\ee
where
\bl
\rho_{k}^{GF} = 3 p_{k}^{GF}
=\frac{k^4}{2\pi^2}
\Big[    \langle\psi|a_{\bf k}^{(3)\dag} a_{\bf k}^{(3)}|\psi\rangle
  -\langle\psi|a_{\bf k}^{(0)\dag} a_{\bf k}^{(0)} |\psi\rangle
+k \Big] .
\el
(It is remarked that the trace of GF part is zero
in the Minkowski spacetime,
unlike the nonzero trace \eqref{vevtrace} in de Sitter space.)
By the GB condition \eqref{LTGBcondition}, the photon part cancels,
and only the vacuum part remains
\be \label{GFenergymk}
\rho^{GF}_{k}
= 3 p^{GF}_{k} = \frac{k^4}{2 \pi^2 }  \, ,
\ee
which has only one divergent $k^4$ term,
corresponding to the dominant UV divergent
terms of \eqref{rhogf97} \eqref{pressfg} in de Sitter space.
The UV divergence of \eqref{GFenergymk} in the Minkowski spacetime
can be removed by normal ordering also,
yielding a zero GF stress tensor.
The expectation values of
all three parts of the stress tensor are independent of $\zeta$,
and the regularized vacuum stress tensor is zero.
Thus, the properties of the stress tensor
of the Maxwell field with the GF  term
in the Minkowski spacetime
are similar to those in de Sitter space.

The above calculations are based on the modes
 \eqref{A1qt}--\eqref{qtpi0aa0} for  $\zeta\ne 1$.
We may  as well use the modes \eqref{Aposminknf}--\eqref{canmoma0}
for  $\zeta\ne -1$,
implement the covariant canonical quantization,
and get the same stress tensor.

\end{document}